\documentclass[preprint,authoryear,12pt]{elsarticle}
\usepackage[centertags]{amsmath}
\usepackage{amsfonts}
\usepackage{amssymb}
\usepackage{amsthm}
\usepackage{epsfig}
\usepackage{setspace}
\usepackage{comment}
\usepackage{ae} 
\usepackage[authoryear]{natbib}

\usepackage[usenames]{color}
\theoremstyle{plain}
\newtheorem{thm}{Theorem}[section]
\newtheorem{lem}[thm]{Lemma}
\newtheorem{cor}[thm]{Corollary}
\newtheorem{prop}[thm]{Proposition}
\theoremstyle{definition}

\theoremstyle{remark}
\newtheorem{exmp}[thm]{Example}
\newtheorem{rem}[thm]{Remark}

\renewcommand{\epsilon}{\varepsilon}
\renewcommand{\P}{\operatorname{P}}

\renewcommand{\i}{\mathrm{i}}

\newcommand {\argmin}{\operatorname{argmin}}

\newcommand{\E}{\operatorname{E}}

\renewcommand{\Re}{\operatorname{Re}}
\renewcommand{\Im}{\operatorname{Im}}

\newcommand{\Cov}{\operatorname{Cov}}
\newcommand{\arginf}{\operatorname{arginf}}

\begin{document}
\begin{frontmatter}
\title{Spectral estimation of the L\'evy density in partially observed affine models}
\author{Denis Belomestny\corref{cor1}\fnref{label}}
\fntext[label]{This work has been supported in part by the Collaborative Research Center
`Economic Risk' (SFB 649) of the German
Research Foundation (DFG).}
\address{Duisburg-Essen University,  Duisburg, Germany}

\begin{abstract}
The problem of  estimating the L\'evy density of a partially observed multidimensional  affine process from   low-frequency and mixed-frequency data   is considered.
The estimation methodology  is based
on the log-affine representation of the conditional characteristic function of an affine
process and local linear smoothing in time.
We derive almost sure uniform  rates of convergence
for  the estimated L\'evy density both in mixed-frequency and low-frequency setups and prove that these rates are optimal in the minimax sense.  Finally, the performance of the estimation  algorithms is illustrated in the case of the Bates stochastic volatility  model.
\end{abstract}

\end{frontmatter}

\section{Introduction}
The problem of  nonparametric statistical inference for jump processes or more generally for semimartingales models has long history and goes back to the works of \citet{RT} and \citet{BB}.  The recent revival of interest in this topic documented, for example, in \citet{FL}, is mainly related to the  wide availability of financial and economical time series data  and new types of statistical issues that have not been addressed before.
For instance, there is now considerable evidence (see, e.g. \citet{CM}) that most financial time series contain a continuous martingale component. This is why in a number of recent  works the problem of estimating some characteristics of jumps for the general semimartingale models with a nonzero continuous part was studied.
In fact, without any further assumptions such kind of statistical inference would not be possible because the behavior of the jump component becomes statistically indistinguishable from the behavior of the diffusion part as the activity of  small jumps increases. In the case of L\'evy processes the activity of small jumps can be measured by the so-called
Blumenthal-Getoor index. The nearer is the Blumenthal-Getoor
index to \( 2 \), the more difficult becomes the problem of separating  jump and diffusion components and hence the problem of statistical inference on the characteristics of jumps (see, e.g. \citet{NR}).
Suppose that the values of a  process \( X(t)\) on a time grid \( \pi=\{ t_{0},t_{1},\ldots, t_{n} \} \) are observed. If \( |\pi| \) is small (high-frequency data), then a large increment \( X(t_{i})-X(t_{i-1}) \)
indicates that a jump occurred between time \( t_{i-1} \) and \( t_{i} \). Based on this
insight and the continuous-time observation analogue, inference for various characteristics of jumps of the underlying semimartingale can be conducted.
For example, in  \citet{AJ2} the problem of statistical inference on the degree of jump activity
in the general semimartingale models based on high-frequency data was considered.
They proposed an estimation procedure which is able to ``see through'' the continuous part  of the semimartingale and consistently estimate the degree of small jump activity  under some restrictions on the structure of the underlying
semimartingale. In fact, these restrictions  keep the highest degree of
activity of small jumps away from \( 2 \), thus allowing for a consistent estimation of the degree of jump activity.
\par
In this paper we focus on a special class of semimartingale models, namely the so-called affine models.
Affine It\^o-L\'evy models are nowadays rather popular in  financial and econometric modeling. Due
to their analytical tractability  on the one hand
and their rather rich dynamics and implied volatility patterns, on the other hand, they
are particularly useful in the context of option pricing.
Many well known models such as Heston and Bates stochastic volatility models fall into the class of affine It\^o-L\'evy models. Option pricing in these models can be conveniently done via the Fourier method.
The literature on affine processes is rather extensive.
 Let us mention two seminal papers of \citet{DPS} and \citet{DFS}, where  theoretical analysis of regular affine models was conducted.
 \par
 In this work we consider the problem of estimating the characteristics of jumps in a class of  affine models with a nonzero continuous part, where it  is assumed that  only  first few  components of the underlying affine process \( X \) are observable at low or mixed-frequency.
We propose an approach based on the log-affine representation of the conditional characteristic function
of an affine  process. This representation  together with some transformation  allows one to consistently
estimate the characteristics of the jump component from  low-frequency and mixed-frequency data
 under some prior bound on the highest degree of activity of small jumps. We present
 uniform convergence rates for the so constructed estimate of
 the transformed L\'evy density which turn out to be optimal in the minimax sense. As the main technical result, that may be of independent interest,
we provide exponential  inequalities on the probability of large deviations for the kernel type empirical processes in uniform metric for the case of weakly dependent random variables.
\par
The problem of parametric estimation of the characteristics of an affine jump-diffusion process
 (processes with finite intensity of jumps)  \( X(t) \) from high-frequency time series of the asset \( S(t)=\exp(X(t))\)  has been recently considered in the literature by \citet{S} and \cite{B2}.
In \citet{S} the general method of moments (GMM) based on the empirical characteristic function was employed and the asymptotic properties  of the corresponding estimator are investigated. \cite{B2} proposed a
filtration-based maximum likelihood methodology for estimating the parameters and the realizations of latent affine processes.
Since the characteristics  of a general affine process are a priori an infinite-dimensional object, any
parametric approach is always exposed to the problem of
misspecification, especially if there is no inherent economic
foundation for the parameters and they are only used to generate
different shapes of possible jump distributions.
The problem of  semi-parametric inference for the characteristics of  special  type affine  processes \( X(t) \) was  studied in the literature as well. In the case of  high-frequency observations,  the problem of  nonparametric
inference on the L\'evy measure of  the time-changed L\'evy processes, belonging sometimes to the
class of affine processes, has been recently studied in \citet{FL1}.
In \citet{JMV} the case of a one-dimensional L\'evy driven Ornstein-Uhlenbeck process, affine process with  zero diffusion part,  was considered. The authors assumed that the corresponding jump component is self-decomposable  and proposed a  cumulant \( M \)-estimator
to estimate the so-called canonical function of the driving self-decomposable process from low-frequency data.
As to the special case of L\'evy processes, semi-parametric estimation  for pure L\'evy
models under low-frequency data  has recently been  studied in \citet{NR}. Let us mention that
in \citet{NR}  the diffusion component is assumed to be known. Thus, all the above works do not encounter the problem of separating  diffusion
and jump components as the activity of small jumps increases. Furthermore, the challenge of devising nonparametric estimation methods for the L\'evy density in general  affine models
lies in the fact that the structure of the conditional characteristic function does not have such explicit form as in the case of pure L\'evy processes and is related to the parameters of the underlying affine process not directly but via a Riccati equation.  The last but not the least: the increments of the general affine process are not  independent, hence advanced tools from the time series analysis have to be used.
\par
The paper is organized as follows. In Section~\ref{main_setup} we introduce the main object of our study, the affine  It\^o-L\'evy
processes and formulate the main existence result. In  Section~\ref{main_ideas} the main ideas behind our estimation methodology are sketched and the notations is introduced. The estimation algorithms for the cases of mixed-frequency and low-frequency data are presented in Section~\ref{rho_est} and Section~\ref{rhotilde}, respectively. The asymptotic properties of the constructed estimates are studied in Section~\ref{asymp}. Section~\ref{NE} contains some numerical examples. The exponential inequalities for the kernel type empirical processes are given in Section~\ref{ExpIneq}. Finally, the proofs of the main results are collected in Section~\ref{proofs}.
\section{Main setup}
\label{main_setup}
Let us fix a probability space
$(\Omega, \mathcal{F},P)$
and an information filtration \( (\mathcal{F}_{t})_{t\ge0} \).
The process \( X(t) \) is an affine process if it is stochastically continuous, time-homogenous
Markov process  with the state space \( \mathcal{D}\subset \mathbb{R}^{d} \), such that
the conditional characteristic function  of \( X(t) \) given \( X(0) \) is an affine function of the initial
state \( X(0) \):
\begin{equation}
\label{phi}
\phi(u|s,x):=\E\left(\left.e^{\mathfrak{i}u^{\top}X(s)}\right| X(0)=x \right)=e^{\psi_{0}(u,s)+x^{\top}\psi_{1}(u,s)},
\quad u\in \mathbb{R}^{d}.
\end{equation}
The affine  process \( X(t) \) is called regular, if the derivatives
\begin{eqnarray*}
    F_{0}(u):=\left.\frac{\partial \psi_{0}(u,s)}{\partial s}\right |_{s=0},
    \quad
    F_{1}(u):=\left.\frac{\partial \psi_{1}(u,s)}{\partial s}\right |_{s=0}
\end{eqnarray*}
exist and are continuous at \( u=0. \) As was recently shown by \citet{KST}, any affine process is, in fact, regular.  The following theorem provides the characterization of affine processes and is proved in \citet{DFS}.
\begin{thm}
\label{CAP}
If \( (X(t))_{t\geq 0} \) is a regular affine process, then the complex valued functions \( \psi_{0} \) and \( \psi_{1} \)
satisfy the following (generalized) Riccati equations
\begin{eqnarray}
\label{FF0}
  \frac{\partial \psi_{0}(u,s)}{\partial s}&=&F_{0}(\psi_{1}(u,s)), \quad  \psi_{0}(u,0)=0,
  \\
\label{FF1}
  \frac{\partial \psi_{1}(u,s)}{\partial s}&=&F_{1}(\psi_{1}(u,s)), \quad  \psi_{1}(u,0)=\i u,
\end{eqnarray}
where
\begin{eqnarray*}
F_{0}(z)&=&(\alpha^{(0)} z,z)+(z,\beta^{(0)})-\gamma^{(0)}+\int_{\mathcal{D}\setminus \{ 0 \}}\left( e^{z^{\top}u}-1-(\chi(u),z) \right)\, \nu^{(0)}(du),
\\
F_{1,j}(z)&=&(\alpha^{(1)}_{j} z,z)+(z,\beta_{j}^{(1)})-\gamma_{j}^{(1)}+\int_{\mathcal{D}\setminus \{ 0 \}}\left( e^{z^{\top}u}
-1-(\chi(u),z) \right)\, \nu^{(1)}_{j}(du)
\end{eqnarray*}
for \( j=1,\ldots,d \) and \( \chi(u)=(\chi_{1}(u),\ldots,\chi_{d}(u)) \) with
\begin{eqnarray*}
  \chi_{k}(u)=
  \begin{cases}
  0, & u_{k}=0,
  \\
  (1\wedge |u_{k}|)\frac{u_{k}}{|u_{k}|}, & \mbox{otherwise}
  \end{cases}
\end{eqnarray*}
for \( k=1,\ldots,d. \) Here \( \alpha=(\alpha^{(0)},\alpha^{(1)})\in
\mathbb{R}^{d\times d}\times\mathbb{R}^{d \times d \times d}, \) \(  \beta=(\beta^{(0)},\beta^{(1)})\in
\mathbb{R}^{d}\times\mathbb{R}^{d \times d}, \)  \(  \gamma=(\gamma^{(0)},\gamma^{(1)})\in
\mathbb{R}\times\mathbb{R}^{d} \)and \( \nu=(\nu^{(0)}, \nu^{(1)}_{1}, \ldots, \nu^{(1)}_{d}) \)
is a vector of measures on \( \mathbb{R}^{d}, \) satisfying
\begin{eqnarray*}
 \int_{\mathcal{D}\setminus\{ 0 \}}\| \chi(u) \|_{2}^{2} \, \nu^{(0)}(du)< \infty, \quad   \int_{\mathcal{D}\setminus\{ 0 \}}\| \chi(u) \|_{2}^{2} \, \nu^{(1)}_{j}(du)< \infty, \quad j=1,\ldots, d,
\end{eqnarray*}
where here and in the sequel \( \| x \|_{2}:=\sqrt{x^{2}_{1}+\ldots+x^{2}_{d}} \) for any \( x\in \mathbb{R}^{d}. \)
\end{thm}
Under some admissibility conditions a regular affine process \( X(t) \)
is a Feller process in the domain \( \mathcal{D}= \mathbb{R}^{m}\times \mathbb{R}_{+}^{d-m}\)  \citep[see][Section 2]{DFS}, where the function \( F_{0} \) corresponds to  the state-independent part of the infinitesimal operator  and \( F_{1} \) is related to the state-dependent one. The admissibility conditions imply, in particular, that
\begin{eqnarray}
\label{admiss1}
  \alpha^{(0)}_{ij}=0, \mbox{ if } i\in \{ m+1,\ldots,d \} \mbox{ or } j\in \{ m+1,\ldots,d \}.
\end{eqnarray}
and \( \alpha^{(1)}_{j}\equiv 0  \) for \( j=1,\ldots, m.  \)
In the sequel we assume that the above admissibility conditions  hold. Moreover, we restrict our analysis to the class of  affine processes with state-independent jumps, i.e.,
\begin{eqnarray}
\label{AsNu1}
\nu^{(1)}_{1}\equiv\ldots\equiv \nu^{(1)}_{d}\equiv 0.
\end{eqnarray}
On the one hand, this assumption reduces the dimensionality of the jump component of \( X \). On the other hand,
the class of affine models satisfying \eqref{AsNu1}, remains rather large and includes such well known
models as Heston, Bates and Barndorff-Nielsen and Shephard  stochastic volatility models. In this paper we study the problem of statistical inference based on partially observed affine processes. In particular, we assume that only the first \( m \) components of the process \( X \) are observed (as it is usual in the case of stochastic volatility models). As a result, in order to ensure identifiability, we have to assume additionally that \( \operatorname{supp}\nu^{(0)}\subset \mathbb{R}^{m}, \) i.e., the positive part of the process \( X \) does not have jumps.
\section{Main ideas}
\label{main_ideas}
Assume that the process \( X(t) \) is stationary with the stationary distribution \( \pi.  \)
Fix some \( x\in \mathbb{R}^{m}, \)  \( s\in \mathbb{R}_{+} \) and denote
\begin{equation}
\label{psi}
 \psi(v|s,x)=\psi_{0}((v,0,\ldots,0),s)+\left(x,\E_{\pi}[ X_{m+1}(0)],\ldots,\E_{\pi} [ X_{d}(0)]\right)^{\top}\psi_{1}((v,0,\ldots,0),s)
\end{equation}
for any \( v\in \mathbb{R}^{m}. \)  Introduce
the function
\begin{eqnarray}
   \label{Transform}
    \Psi(v|s,x):=\int_{[-1,1]^{m}} \left(\psi(v|s,x)-\psi(v+w|s,x) \right) \, dw.
\end{eqnarray}
Let us now investigate  the behavior of the function \( \Psi \) as \( s\to 0 \) (``short time asymptotic'') and \( s\to \infty \) (``long time asymptotic'').
\paragraph{Short time asymptotic}
Due to \eqref{FF0} and \eqref{FF1}, it holds
\begin{eqnarray}
\label{PsiRepr1}
    \Psi_{0}(v)=\left.\frac{\partial\Psi(v|s,x)}{\partial s}\right|_{s=0}=
    \mathcal{L}+\int_{\mathbb{R}^{m}}e^{\i z^{\top}v} \rho^{(0)} (dz)
\end{eqnarray}
 for some \( \mathcal{L}\in \mathbb{R} \) depending linearly on \( x, \) where
\begin{eqnarray}
   \label{rho_nu}
    \rho^{(0)}(dz)&:=&2^{m}\prod_{k=1}^{m} \left( 1-\frac{\sin z_{k}}{z_{k}} \right)\nu^{(0)}(dz)
\end{eqnarray}
is a finite measure on \( \mathbb{R}^{m} \).
\paragraph{Long time asymptotic} If the maximal eigenvalue of the matrix \( (\beta^{(1)}_{ij})_{1\leq i,j \leq d} \) is negative, then the affine process \( X \) is ergodic, possesses a stationary distribution \( \pi \) (see \citet{MH}) and it holds \( \phi(v,0,\ldots,0|s,x)\to \phi_{\pi}(v) \) as \( s\to +\infty, \) with
\begin{eqnarray*}
    \phi_{\pi}(v)=\exp\left( \int_{0}^{\infty}F_{0}(\psi_{1}((v,0\ldots,0),s))\,ds \right), \quad v\in \mathbb{R}^{m}.
\end{eqnarray*}
If
\begin{eqnarray}
\label{AssBeta0}
\beta^{(0)}_{m+1}=\ldots=\beta^{(0)}_{d}=0,
\end{eqnarray}
then the admissibility condition \eqref{admiss1} implies
\begin{multline}
\label{PsiRepr2}
    \psi_{\pi}(v)=\log \{ \phi_{\pi}(v) \}=-(v,\widetilde\alpha^{(0)}v)+\i(\widetilde\beta^{(0)},v)
    \\
    +\int_{0}^{\infty}\int_{\mathbb{R}^{m}}\left( e^{\i( e^{s\mathfrak{B}}v,w)}-1-\i(\chi(w),e^{s\mathfrak{B}}v) \right)\, \nu^{(0)}(dw)\, ds
\end{multline}
for some  \( \widetilde\alpha^{(0)}\in \mathbb{R}^{m\times m}, \) \(\widetilde\beta^{(0)}\in \mathbb{R}^{m} \) and \( \mathfrak{B}=(\beta^{(1)}_{ij})_{1\leq i,j \leq m}.  \)
Therefore
\begin{eqnarray}
\label{PsiRepr}
    \Psi_{\pi}(v)=\lim_{s\to+\infty}\Psi(v|s,x)=
    \widetilde{\mathcal{L}}+\int_{\mathbb{R}^{m}}e^{\i z^{\top}v} \widetilde\rho^{(0)} (dz)
\end{eqnarray}
 with some \( \widetilde{\mathcal{L}}>0 \) and
\begin{eqnarray}
   \label{rhotilde_nu}
    \widetilde\rho^{(0)}(dz)&:=&2^{m}\prod_{k=1}^{m} \left( 1-\frac{\sin z_{k}}{z_{k}} \right)\widetilde\nu^{(0)}(dz),
\end{eqnarray}
where for any set \( A\subset \mathcal{B}(\mathbb{R}^{m}) \)
\begin{eqnarray*}
    \widetilde\nu^{(0)}(A)=\int_{0}^{\infty}\int_{\exp(s\mathfrak{B}^{*})q\in A} \nu^{(0)}(dq)\, ds.
\end{eqnarray*}
In the sequel we shall assume that the measure \( \nu^{(0)}  \) is absolutely continuous w.r.t. Lebesgue measure on \( \mathbb{R}^{m} \) and have a bounded density. Then \( \widetilde \nu^{(0)} \) has also a density given by
\begin{eqnarray*}
\widetilde \nu^{(0)}(z)=\int_{0}^{\infty}|e^{-s\mathfrak{B}^{*}}|\nu^{(0)}(e^{-s\mathfrak{B}^{*}}z)\, ds
\end{eqnarray*}
and the admissibility conditions imply that
\begin{eqnarray*}
    \max_{k=1,\ldots,m}\inf\left\{r\geq 0:\int_{\{|x_{1}|>1,\ldots,|x_{k-1}|>1,\,  |x_{k}|\leq 1, \, |x_{k+1}|>1,\ldots, |x_{m}|>1 \}} |x_{k}|^{r}\nu^{(0)}(dx)<\infty \right\}<1.
\end{eqnarray*}
Hence, the measures  \( \rho^{(0)}  \) and \( \widetilde\rho^{(0)}  \) are absolutely continuous
w.r.t. Lebesgue measure on \( \mathbb{R}^{m} \) as well and possess bounded densities denoted (with some abuse of notations) by \( \rho^{(0)}(x) \) and \( \widetilde\rho^{(0)}(x), \) respectively.
Functions \(\Psi_{0}(v) \) and \( \Psi_{\pi}(v) \) satisfy, due to the Riemann-Lebesgue theorem, the following asymptotic relations
\begin{eqnarray*}
\lim_{\| v \|\to \infty}  \Psi_{0}(v)&=&\mathcal{L},
  \\
\lim_{\| v \|\to \infty}  \Psi_{\pi}(v)&=&\widetilde{\mathcal{L}}.
\end{eqnarray*}
The above  relations together with \eqref{PsiRepr1} and \eqref{PsiRepr2} indicate that  one can estimate the Fourier transforms of \( \rho^{(0)}(x) \) and \( \widetilde\rho^{(0)}(x), \) if some estimates for the functions \(\Psi_{0}(u) \) and \( \Psi_{\pi}(u) \) are available. In order to estimate
\( \Psi_{0} \) using formula \eqref{PsiRepr1} we need to perform   numerical differentiation of the function \( \Psi(v|s,x) \) in \( s. \) This calls for the ``high-frequency'' data. On the other hand, in order to estimate the expectation in \( \Psi \) by  a kind of averaging, we need ergodic theorem which usually holds only under low-frequency sampling. It turns out that by mixing low- and high-frequency data one can consistently estimate \( \Psi_{0}(u). \) As to the function \( \Psi_{\pi}, \) it can be estimated from low-frequency data.
\begin{rem}
In the case of the more general affine processes, i.e., in the case where \( \nu^{(1)}_{k}\not\equiv 0 \) for
some \( k\in \{ 1,\ldots,m \}, \) one can use a similar strategy to reconstruct  \( \nu^{(1)}_{k}, \) \( k=1,\ldots, m. \)
 Indeed, in the general case the function \( \Psi_{0} \) takes the form
\begin{eqnarray}
\label{PsiRepr}
    \Psi_{0}(v)=\mathcal{L}+\int_{\mathbb{R}^{m}}e^{\i z^{\top}v} \rho^{(0)} (dz)+\sum_{j=1}^{m}x_{j}\int_{\mathbb{R}^{m}}e^{\i z^{\top}v} \rho^{(1)}_{j} (dz),
\end{eqnarray}
where each measure \( \rho^{(1)}_{j} (dz) \) is related to \( \nu^{(1)}_{j} (dz) \) in the same way as
\( \rho^{(0)} (dz) \) was related to \( \nu^{(0)} (dz). \)
Therefore, one can reconstruct the Fourier transforms of all measures \( \rho^{(1)}_{j} (dz), \) \( j=1,\ldots,m, \) if one is able to recover function \( \Psi(u|s,x) \) and its derivative in \( s \) for (at most) \( m \) linearly independent vectors \( x. \) In principle, the approach presented in the next section allows one to estimate
 \( \Psi(v|s,x) \) for arbitrary number of vectors \( x. \)
\end{rem}

\section{Estimation of \( \rho^{(0)}(z) \) in the mixed-frequency setup}
\label{rho_est}
\subsection{Observations}
For our theoretical study we adopt the  observational model based on the mixed-frequency  random sampling. In particular, we  assume that a trajectory of a partially observed process \( X \) containing the pairs
\begin{eqnarray*}
( X^{(m)}(t_{j}),X^{(m)}(t_{j}+\delta_{j}) ), \quad j=1,\ldots,n,
\end{eqnarray*}
 is observable, where \( X^{(m)}(t)=(X_{1}(t),\ldots,X_{m}(t)), \) \( \delta _{j}, \, j=1,\ldots, n, \) are i.i.d. random variables  on \( [0,T] \) for some fixed \( T>0 \) with a common  density \( p_{\delta }(x) \) and \( \min_{j}(t_{j+1}-t_{j})>T \). The latter assumption implies that the time horizon \( T+t_{n} \) of observations
 tends to infinity as \( n\to \infty. \) In the sequel (see Assumption (AG) in Section and Remark~\ref{AGR}) we will  additionally assume that the density of the r.v. \( (\delta_{j})  \) does not vanish in the vicinity of \( 0, \) meaning that the r.v. \( (\delta_{j})  \) can, with  positive probability, take values that are arbitrary close to \( 0. \) This is the reason why we call the above observation's scheme mixed-frequency setup. Mixing low-frequency and high-frequency data has become rather popular technique in recent years. It has been, for example, used to improve the volatility estimation  (see, e.g., \citet{TWYZ})  or to achieve better forecasts  in macro-economical models (see, e.g. \citet{GAK} and references therein).  Let us also note that the high-frequency data is usually not equidistant: the observations are more frequent for busy trading days. In such a situation our random sampling scheme may be  appropriate.
 As we will see, the above sampling scheme will also allow us to consistently estimate the function \( \Psi_{0} \) in \eqref{PsiRepr1}. Indeed, while the high-frequency sampling scheme (small values of \( (\delta_{j})\)) makes it possible to estimate the derivative of the function \( \Psi(v|s,x) \) in \( s \), the low-frequency data  allows us to consistently estimate the function \( \Psi(v|s,x) \) by its empirical counterpart. The condition
\( t_{j+1}-t_{j}-\delta_{j}>0 \) ensures that the subsequent pairs are only weakly dependent and a kind of ergodic theorem can be applied.
 \subsection{Estimation of \( \Psi_{0}(v) \)}
 \label{Psi0}
Assume again that the process \( X(t) \) is stationary with the stationary distribution \( \pi.  \) In this section we shall, for any fixed \( x\in \mathbb{R}^{m}, \) estimate the quantity
\begin{eqnarray}
 \label{phisex}
 \left. \partial\phi\left((v,0\ldots,0)|s,\left(x, \E_{\pi } [X_{m+1}(0)],\ldots,\E_{\pi } [X_{d}(0)] \right)\right)\partial s\right |_{s=0},
\end{eqnarray}
by local polynomial
smoothing (local polynomial of degree \( l \) in \( x \)  and local linear in \( s \)) of the empirical characteristic process
\begin{eqnarray*}
     Z_{j}(v)=\exp(\i v^{\top} X^{(m)}(t_{j}+\delta_{j})), \quad v\in \mathbb{R}^{m}, \quad j=1,\ldots,n.
\end{eqnarray*}
Since the process \( X(t) \) is stationary, \eqref{phisex} is equal to
\begin{eqnarray*}
    \E_{\pi(X_{m+1},\ldots, X_{d})}\left[ \left. \partial\phi\left((v,0\ldots,0)|s,(x, X_{m+1},\ldots, X_{d} )\right)/\partial s\right |_{s=0} \right].
\end{eqnarray*}
The latter quantity can be estimated by performing the local polynomial smoothing w.r.t. the first \( m \) components of the process \( X(t) \) and averaging w.r.t to the conditional distribution of the remaining coordinate processes \( X_{m+1}(t),\ldots, X_{d}(t). \) This is basically what we do next.
Fix some \( x\in \mathbb{R}^{m}. \)
For some \( h_{1}>0, \) \( h_{2}>0 \), an integer \( l\geq 0 \) and a function \( K:\mathbb{R}^{1+m}\to \mathbb{R}_{+} \), let \( (\mathcal{Q}_{0,n},\mathcal{Q}_{1,n}) \) be a solution of the following optimization problem
\begin{equation}
    \label{LPO}
    \min_{(\mathcal{Q}_{0},\mathcal{Q}_{1})}\left\{ \sum_{j=1}^{n}w_{j}\left[Z_j(v)-\mathcal{Q}_{0}(X^{(m)}(t_{j})-x)-\delta _{j}\mathcal{Q}_{1}(X^{(m)}(t_{j})-x)\right]^{2}\right\}
\end{equation}
with \( w_{j}=K\left(\delta_{j}/h_{1}, (X^{(m)}(t_{j})-x)/h_{2}\right), \) where the minimization  is performed over  the set of all polynomials \( \mathcal{Q}_{0} \) and \( \mathcal{Q}_{1} \) on \( \mathbb{R}^{m} \)  of degree \( l. \)
Now define the local polynomial estimates for  \( \exp(\psi(v|0,x)) \) and \( \partial_{s}\psi(v|0,x),\) where \( \psi(v|0,x) \) is defined in \eqref{psi}, by
\begin{eqnarray*}
     \phi_{n}(v)=\mathcal{Q}_{0,n}(0), \quad \phi_{s,n}(v)=\mathcal{Q}_{1,n}(0),
\end{eqnarray*}
respectively.
Furthermore, define an estimate for \( \Psi_{0}(v) \) by plugging the estimate \( \phi_{s,n} \) into \eqref{Transform}:
\begin{eqnarray*}
  \Psi_{0,n}(v)&:=&\int_{[-1,1]^{m}} \left(\phi_{s,n}(v)-\phi_{s,n}(v+w)\right) \, dw.
\end{eqnarray*}
Let \( \pi_{0,\mathbf{m}} \) and \( \pi_{1,\mathbf{m}} \) denote the coefficients of the polynomials \( \mathcal{Q}_{0,n} \)
and \( \mathcal{Q}_{1,n}, \)
respectively, indexed by
the multi-index \( \mathbf{m}\in \mathbb{N}^{m} \), i.e., \(\mathcal{Q}_{k,n}(z)=\sum_{|\mathbf{m}|\leq l}\pi_{k,\mathbf{m}}z^{\mathbf{m}}, \) \( k=0,1. \) Introduce the vectors
\( \Pi_{k}=(\pi_{k,\mathbf{m}})_{|\mathbf{m}|\leq l} \) and \( S_{k}=(S_{k,\mathbf{m}})_{|\mathbf{m}|\leq l}, \) \( k=0,1, \) with
\begin{eqnarray*}
    S_{k,\mathbf{m}}=\frac{1}{nh_{1}h_{2}^{m}}\sum_{j=1}^{n}Z_{j}(v)\left(\frac{\delta _{j}}{h_{1}}\right)^{k}\left(\frac{X^{(m)}(t_{j})-x}{h_{2}}\right)^{\mathbf{m}}w_{j}.
\end{eqnarray*}
Let \( P(z)=(z^{\mathbf{m}})_{|\mathbf{m}|\leq l}\) be the vector of all monomials in \( \mathbb{R}^{m} \) of order less
than or equal to \( l \) and the matrices
\( \Gamma_{k}=(\Gamma_{k,\mathbf{m}_{1},\mathbf{m}_{2}})_{|\mathbf{m}_{1}|,|\mathbf{m}_{2}|\leq l}, \, k=0,1,2, \) be defined as
\begin{equation}
\label{Gamma}
  \Gamma_{k,\mathbf{m}_{1},\mathbf{m}_{2}}=\frac{1}{nh_{1}h_{2}^{m}}\sum_{j=1}^{n}\left(\frac{\delta _{j}}{h_{1}}\right)^{k}\left(\frac{X^{(m)}(t_{j})-x}{h_{2}}\right)^{\mathbf{m}_{1}+\mathbf{m}_{2}}w_{j}.
\end{equation}
Consider now the vector \( S=(S_{0},S_{1})^{\top} \) and  the matrix
\begin{eqnarray*}
    \Gamma=
    \begin{pmatrix}
    \Gamma_{0} & \Gamma_{1}\\
    \Gamma_{1} & \Gamma_{2}
    \end{pmatrix}.
\end{eqnarray*}
The following proposition holds.
\begin{prop}
\label{LPR}
If the matrix \( \Gamma \) is positive definite, then there exist  unique polynomials \( \mathcal{Q}_{0,n} \) and \( \mathcal{Q}_{1,n} \) on \( \mathbb{R}^{m} \) of degree \( l \) solving \eqref{LPO}. Their vectors of coefficients are given by
\( \Pi=(\Pi_{0},\Pi_{1})=\Gamma^{-1}S. \)  As a result
\begin{eqnarray}
\label{LPE0}
    (\phi_{n}(v),\phi_{s,n}(v))^{\top}=(P(0),P(0))^{\top}\Gamma^{-1} S.
\end{eqnarray}
\end{prop}
Proposition \ref{LPR} implies that \( \phi_{s,n}(v)=\sum_{j=1}^{n}\omega_{j} Z_{j}(v)\)
with some weights \( \omega_{j}, \) \( j=1,\ldots,n. \) Hence, the following representation for
\( \Psi_{0,n} \) holds
\begin{eqnarray*}
  \Psi_{0,n}(v)=\sum_{j=1}^{n}\bar\omega_{j}Z_{j}(v), \quad v\in \mathbb{R}^{m},
\end{eqnarray*}
where \( \bar\omega_{j}=\omega_{j}\int_{[-1,1]^{m}} (1-\exp(\i v^{\top}X^{(m)}(t_{j})))\, dv. \)
\subsection{Estimation of \( \rho^{(0)}(z) \)}
Let \( \mathcal{K}_{\mathcal{L}}(u) \) be a  regularizing kernel supported on \( [-1,1]^{m} \) and let \( U_{n} \) be a sequence of positive numbers tending to \( \infty. \)
Define  an estimate for the limit \( \mathcal{L} \) in \eqref{PsiRepr1}  as
\begin{eqnarray*}
  \mathcal{L}_{n}&=&\int_{\mathbb{R}^{m}}U^{-m}_{n}\mathcal{K}_{\mathcal{L}}(v/U_{n})\Psi_{0,n}(v)\, du.
\end{eqnarray*}
Next, we reconstruct  the L\'evy density \( \rho^{(0)}(z)\) using the regularized Fourier inversion formula
\begin{eqnarray*}
\label{rho}
\rho^{(0)}_{n}(z)&=&\frac{1}{(2\pi)^{m}}\int_{\mathbb{R}^{m}}e^{-\i v^{\top}x}\mathcal{K}_{\rho }(v/U_{n})\left[ \Psi_{0,n}(v)-\mathcal{L}_{n} \right]\, du,
\end{eqnarray*}
where \( \mathcal{K}_{\rho} \) is another regularizing kernel.
\section{Estimation of \( \widetilde\rho^{(0)}  \)  using low-frequency data}
\label{rhotilde}
\subsection{Observations}
We assume that the time series \( X^{(m)}(t_{1}),\ldots X^{(m)}(t_{n}) \) is observed, where
\( t_{k},\, k=1,\ldots,n, \) is a deterministic sequence of positive numbers with \( t_{k}-t_{k-1}>\Delta, \)  \( k=1,\ldots, n, \) for some \( \Delta>0. \)
\subsection{Estimation of \( \Psi_{\pi } \)}
By Birkhoff ergodic theorem it holds for any \(v\in \mathbb{R}^{m} \)
\begin{eqnarray*}
    \phi_{\pi,n}(v)=\frac{1}{n}\sum_{k=1}^{n} \exp(v^{\top}X^{(m)}(t_{k}))\to \phi_{\pi }(v), \quad n\to \infty,
\end{eqnarray*}
where \( \phi_{\pi}(v) \) stands for the  c.f. of the first \( m \) components of \( X(t) \) under \( \pi.  \)
Therefore it is natural to estimate \( \Psi_{\pi} \) via
\begin{eqnarray*}
\Psi_{\pi,n}(v):=\int_{[-1,1]^{m}}\left[ \log \{ \phi_{\pi,n}(v)\}-\log \{ \phi_{\pi,n}(v+w)\}  \right]\, dw.
\end{eqnarray*}
\subsection{Estimation of \( \widetilde\rho^{(0)}  \)}
We again  first  estimate  the limit \( \widetilde{\mathcal{L}} \)  by
\begin{eqnarray*}
  \widetilde{\mathcal{L}}_{n}&=&\int_{\mathbb{R}^{m}}U^{-m}_{n}\mathcal{K}_{\mathcal{L}}(v/U_{n})\Psi_{\pi,n}(v)\, du.
\end{eqnarray*}
Then the  transformed density \( \widetilde\rho^{(0)}(z)\) can be reconstructed using the regularized Fourier inversion formula
\begin{eqnarray*}
\label{rho}
\widetilde\rho^{(0)}_{n}(z)&=&\frac{1}{(2\pi)^{m}}\int_{\mathbb{R}^{m}}e^{-\i v^{\top}z}\mathcal{K}_{\rho}(v/U_{n})\left[ \Psi_{\pi,n}(v)-\widetilde{\mathcal{L}}_{n} \right]\, dv.
\end{eqnarray*}
\section{Asymptotic analysis }
\label{asymp}
In this section we study the asymptotic properties of the estimates \( \rho^{(0)}  \)
and \( \widetilde\rho^{(0)}.  \)
\subsection{Assumptions}

We need the following assumptions.
\begin{description}
 \item[(AX)] The sequence \( X(t_{k}), \) \( k\in \mathbb{N}, \)  is  strongly mixing with
  the mixing coefficients \( \alpha_{X}  \)  satisfying
\begin{eqnarray*}
    \alpha_{X}(k)\leq \bar\alpha_{0} \exp(- \bar\alpha_{1}k), \quad k\geq 1,
\end{eqnarray*}
  for some \( \bar\alpha_{0}>0  \) and \(  \bar\alpha_{1}>0. \)
 \item[(AN)] The L\'evy measure  \( \nu^{(0)} \) satisfies for some \( p>2 \)
\begin{eqnarray*}
    \int_{\{ \|x\|>1 \}}\|x\|^{p}\nu^{(0)}(dx)<\infty,
\end{eqnarray*}
where here and in the sequel \( \| \cdot \| \)  stands for \( l_{\infty} \) norm.
\end{description}
Let \( p_{m,\pi} \) be the density of  \( X^{(m)}(t) \) under \( \pi.  \)
For any fixed \( s \) and \( x, \) consider a matrix
\begin{eqnarray*}
    \bar \Gamma=
    \begin{pmatrix}
    \bar\Gamma_{0} & \bar\Gamma_{1}\\
    \bar\Gamma_{1} & \bar\Gamma_{2}
    \end{pmatrix},
\end{eqnarray*}
where
\( \bar\Gamma_{k}=(\bar\Gamma_{k,\mathbf{m}_{1},\mathbf{m}_{2}})_{|\mathbf{m}_{1}|,|\mathbf{m}_{2}|
\leq l} \)
are matrices with the elements
\begin{eqnarray*}
    \bar \Gamma_{k,\mathbf{m}_{1},\mathbf{m}_{2}}=\int_{\mathbb{R}^{m}}\int_{0}^{\infty} t^{k}z^{\mathbf{m}_{1}+\mathbf{m}_{2}}
K(t,z)p_{\delta}(h_{1}t) p_{m,\pi}(x+h_{2}z) \,dt\,dz.
\end{eqnarray*}
Note that \( \E\Gamma=\bar \Gamma. \)
We make the following assumption about \( \bar\Gamma.  \)
\begin{description}
  \item[(AG)]  The minimal eigenvalue of the matrix \(  \bar\Gamma \) is bounded away from zero, i.e.,
  \begin{eqnarray*}
   \label{EV_WGW}
   \min_{\| W \|=1}\left[ W^{\top}\bar\Gamma W \right]\geq\gamma_{0}
  \end{eqnarray*}
  with some \( \gamma_{0}>0.  \)
   \item[(AP)] The density \( p_{m,\pi} \) is uniformly bounded on \( \mathbb{R}^{m}. \)
   \item[(AK1)] Kernel \( K \) is bounded and is  supported on \( [0,1]\times [-1,1]^{m} \).
   \item [(AK2)] The regularizing kernels \( \mathcal{K}_{\mathcal{L}} \) and \( \mathcal{K}_{\rho} \)  are uniformly bounded, are supported on \( [-1,1]^{m}, \) integrate to \( 1 \)   and satisfy
\begin{eqnarray*}
    \mathcal{K}_{\mathcal{L}}(u)=0, \quad \mathcal{K}_{\rho}(u)=1,\quad u\in [-a_{K},a_{K}]^{m}
\end{eqnarray*}
with some \( 0<a_{K}<1. \)
\paragraph{Discussion}
Exponentially strongly mixing holds for a wide class  of It\^o-L\'evy processes.
In \citet{MH} conditions are formulated that ensure that a multidimensional It\^o-L\'evy process  is  exponentially \( \beta  \)-mixing
 and hence exponentially \( \alpha  \)-mixing.
\begin{exmp}
Let  \( A \)
be \( d\times d  \) matrix whose eigenvalues have
positive real parts, and let \( Z \) be a nontrivial \( d \)-dimensional L\'evy process. Consider  a \( d \)-dimensional Ornstein-Uhlenbeck process  \( X \) given by
\begin{eqnarray*}
    dX(t)=-AX(t)dt+dZ(t), \quad X_{0}\sim \eta.
\end{eqnarray*}
The process \( X \) is obviously an affine process satisfying \eqref{AsNu1}. If the L\'evy measure \( \nu \) of the process \( Z \) satisfies  \( \int_{\|x\|>1}\|x\|^{q}\nu(dx)<\infty \) and \( \int_{\|x\|>1}\|x\|^{q}\eta(dx)<\infty \) for some \( q>0, \) then \( X \) is exponentially \( \beta \)-mixing (see \citet{MH}, Theorem 2.6).
\end{exmp}
Suppose now that \( X \) is an affine process with the characteristics \( \chi \) satisfying admissibility conditions, the condition \eqref{AsNu1} and \( \int_{\|x\|>1}\|x\|^{q}\nu^{(0)}(dx)<\infty \) for some \( q>0 \).
If the maximal eigenvalue of the matrix \( \beta^{(1)}  \) is negative, then both sequences \( X(t_{n}) \)
and \( X(t_{n}+\delta_{n}) \) are exponentially
\( \beta  \)-mixing and hence ergodic.
\begin{rem}
\label{AGR}
Let us remark on the assumption (AG). If for any \( R>0 \) the joint density \( p_{\delta}(t)p_{m,\pi}(x) \) is strictly positive on \( (0,T]\times \mathcal{B}_{R}, \) where \( \mathcal{B}_{R} \) is a  ball of radius \( R \) in \( \mathbb{R}^{m} \), then (AG) holds. Suppose, for simplicity, that \( p_{m,\pi}(x) \) is supported on \( \mathcal{B}_{R} \) for some \( R>0 \) (otherwise a truncation argument combined with an assumption on the tails of \( p_{m,\pi } \) can be used) and consider the kernel
\[
K(t,z):=\frac{\Gamma(1+(m+1)/2)}{\pi ^{(m+1)/2}}\mathbf{1}_{\left\{ \sqrt{t^{2}+\|z\|^{2}}\leq1 \right\}} .
\]
We have for any  \( W=(W_{1},W_{2})\in \mathbb{R}^{D}\times \mathbb{R}^{D} \) with \( D=m(m+2)\cdot\ldots\cdot(m+l-1)/l ! \)
\begin{eqnarray*}
  W^{\top}\bar\Gamma(x) W&=&\int_{0}^{T}\int_{\mathbb{R}^{m}}\left( \sum_{|\alpha|\leq l} W_{1}^{\alpha }z_{\alpha } +t\sum_{|\alpha|\leq l} W_{2}^{\alpha }z_{\alpha }\right)^{2} K(t,z)p_{\delta}(h_{1}t) p_{m,\pi}(x+h_{2}z)\,dt\, dz
  \\
  &\geq&
  B\int_{\mathcal{S}(x,R,T)}\left( \sum_{|\alpha|\leq l} W_{1}^{\alpha }z_{\alpha } +t\sum_{|\alpha|\leq l} W_{2}^{\alpha }z_{\alpha }\right)^{2}\, dt\, dz
\end{eqnarray*}
with some positive constant \( B\) and
\begin{eqnarray*}
\mathcal{S}(x,R,T):=\{ (t,z)\in \mathbb{R}_{+}\times \mathbb{R}^{m}: t^{2}+\| z \|^{2}\leq 1,\, \| x+h_{2}z\|^{2}+h_{1}^{2}t^{2}\leq R^{2}+T^{2} \}.
\end{eqnarray*}
Using now the fact that the Lebesgue  measure of the set \( \mathcal{S}(x,R,T) \)  is larger than some positive number \( \lambda \) for all \( x\in \mathcal{B}_{R}, \) where \( \lambda  \) depends on \( R,T \) and \( d \) but does not depend on \( h_{1} \) and \( h_{2}, \) we get
\begin{eqnarray*}
       \inf_{x\in \mathcal{B}_{R}}\left[ W^{\top}\bar\Gamma(x) W \right]\geq B\inf_{\| W \|=1}\inf_{\mathcal{S}: |\mathcal{S}|>\lambda}\int_{\mathcal{S}} \left( \sum_{|\alpha|\leq l} W_{1}^{\alpha }z_{\alpha } +t\sum_{|\alpha|\leq l} W_{2}^{\alpha }z_{\alpha }\right)^{2} \, dt\, dz\geq \gamma_{0}
\end{eqnarray*}
with some positive \( \gamma_{0} \) by the compactness argument.
\end{rem}
\end{description}
\subsection{Minimax upper bounds for \(  \rho^{(0)}_{n}  \)}
First, introduce a class of L\'evy densities for which we are going to derive the minimax rates of convergence. For  any \( 1\leq\varkappa<2  \)  let \( \mathfrak{L}_\varkappa\) stand for a class of L\'evy densities \( \nu^{(0)}  \) satisfying
\begin{eqnarray}
\label{ass_nu}
\left( \prod_{k=1}^{d}v_{k} \right)^{\varkappa  }\mathcal{F} [\rho^{(0)}] (v)\leq C, \quad v\in \mathbb{R}^{m}
\end{eqnarray}
for some \( C>0, \) where the function \( \rho^{(0)}\) is related to the L\'evy density \( \nu^{(0)} \) via \eqref{rho_nu}.
Here and in the sequel \( \mathcal{F} [\rho] (u) \) stands for the Fourier transform of a measure \( \rho  \).
\begin{rem}
It can be shown that if the L\'evy measure \( \nu^{(0)} \)  satisfies
\begin{eqnarray*}
    \max_{k=1,\ldots,m}\inf\left\{r\geq 0:\int_{\{|x_{1}|>1,\ldots,|x_{k-1}|>1,\,  |x_{k}|\leq 1, \, |x_{k+1}|>1,\ldots, |x_{m}|>1 \}} |x_{k}|^{r}\nu^{(0)}(dx)<\infty \right\}<2-\varkappa,
\end{eqnarray*}
i.e., the degree of jump activity (for the definition see \citet{AJ2}) of  the each component of the process \( X(t) \) is less than \( 2-\varkappa  \), then the inequality \eqref{ass_nu} holds with some \( C>0 \).
\end{rem}
The first result of this section concerns the asymptotic properties of the estimate  \( \phi_{s,n}(v) \) constructed in Section~\ref{Psi0}.
\begin{thm}
\label{phi_conv}
Suppose that the assumption (AX), (AN), (AG), (AP) and (AK1) hold.
Let  \( \phi_{n,s}(v) \) be the local polynomial estimate of degree \( l \) (in \( x \)) for the function  \( \phi_{s}(v)=\partial_{s}\psi(v|0,x),\) where \( \psi(v|0,x) \) is defined in \eqref{psi}.   Furthermore, let $w$ be a monotone positive Lipschitz function on $\mathbb{R}_{+}$ such that
\begin{equation}
\label{decreasing_w}
0<w(z)\leq 1/(1+z)^{2(l+1)}, \quad z\in \mathbb{R}_{+}.
\end{equation}
Then under the choices \( h_{1}= \left(n^{-1}\log^{1+r} n\right)^{(l+1)/(2m+5(l+1))}\) and \( h_{2}=\left(n^{-1}\log^{1+r} n\right)^{1/(m+5(l+1)/2)} \) with some \( r>0, \)
\begin{eqnarray}
\P\left( \sup_{v\in \mathbb{R}^{m}}
\left[ w(\| v \|)|\phi_{n,s}(v)-\phi_{s}(v)| \right]>A\zeta_{n} \right)\leq Bn^{-1-\delta},
\end{eqnarray}
where
\begin{eqnarray}
\label{zeta1}
    \zeta_{n}&=&\left[n^{-1}\log^{(1+r)} n  \right]^{\frac{1+l}{m+5(1+l)}}
\end{eqnarray}
and \( \delta, \) \( A \)  and \( B \) are some positive constants.
\end{thm}
\begin{rem}
The condition \eqref{decreasing_w} on the decay of the weighting function \( w \) can not be, in general, weakened.
For example, in the case of a one-dimensional Brownian motion with volatility \( \sigma^{2}  \) starting at \( 0 \), the simplest affine process, we get
\begin{eqnarray*}
\partial_{ss}\phi(u|s,x)|_{s=0}=4^{-1}\sigma^{4}u^{4}.
\end{eqnarray*}
This means that the approximation errors
of the estimates  \( \phi_{n}(u) \) and \( \phi_{s,n}(u) \)  based on local constant smoothing in \( x \) (\( l=0 \)),  are of order \( (h_{1}^{2}+h_{2}^{1})\sigma^{4} u^{4}/8\)  and
\( (h_{1}+h_{1}^{-1}h_{2}^{1})\sigma^{4} u^{4}/8,\) respectively. So in order to be able to prove the uniform consistency in \( u \) we have to assume
\eqref{decreasing_w}. In fact, the rate \eqref{zeta1} can be proved  to be optimal, provided the function \( \psi(v|s,x) \) is at least two times differentiable in \( s \)
and all partial derivatives in \( x \) up to order \( l+1 \) exist (see \citet{St} for lower bounds for local polynomial estimates).
\end{rem}
Let \( r_{n}  \) be a sequence of positive r.v. and \( q_{n} \) be a sequence of positive real numbers.
We shall write \( r_{n}=O_{a.s.}(q_{n}) \) if there is a constant \( D>0 \) such that \( \P(\limsup_{n\to \infty} q_{n}^{-1}r_{n}\leq D)=1.  \) In the case \( \P(\limsup_{n\to \infty} q_{n}^{-1}r_{n}=0)=1 \) we shall write \( r_{n}=o_{a.s.}(q_{n}). \) Theorem~\ref{phi_conv} implies the following result on the strong uniform rates of convergence for the estimate \( \rho^{(0)}_{n}. \)
\begin{thm}
\label{UpperBounds1}
 Suppose that the assumptions (AX), (AN), (AG), (AP), (AK1) and (AK2)  hold.
Let \( \rho^{(0)}_{n} \) be the estimate for the transformed L\'evy density \(\rho^{(0)}\) defined in Section~\ref{rho_est}.
If \( \nu^{(0)}\in \mathfrak{L}_\varkappa \)
for some \( 1\leq \varkappa<2, \)  then
\begin{eqnarray*}
     \left\|\rho^{(0)}-\rho^{(0)}_{n}\right\|_{L_{\infty}(\mathbb{R}^{m})}=O_{a.s.}\left( \zeta_{n} \int_{[-U_{n},U_{n}]^{m}}w^{-1}(\| v \|)\,dv+U_{n}^{-(\varkappa-1)}\right),
\end{eqnarray*}
with \( \zeta_{n} \) being defined in \eqref{zeta1}.
\end{thm}
\subsection{Minimax upper bounds for \( \widetilde \rho^{(0)}_{n}  \)}
 For  any \( 1\leq\varkappa<2  \)  let \( \widetilde{\mathfrak{L}}_\varkappa\) stand for a class of L\'evy densities \( \nu^{(0)}  \) satisfying
\begin{eqnarray}
\label{ass_nu}
\left( \prod_{k=1}^{d}v_{k} \right)^{\varkappa  }\mathcal{F} [\widetilde\rho^{(0)}] (v)\leq C, \quad v\in \mathbb{R}^{m}
\end{eqnarray}
for some \( C>0, \) where the function \( \widetilde\rho^{(0)}\) is related to the L\'evy density \( \widetilde\nu^{(0)} \) via \eqref{rho_nu}.
We need the following assumption concerning the asymptotic behavior of the sequence \( U_{n} \).
\begin{description}
  \item [(AH)] The sequence  \( U_{n} \) satisfies
  \begin{eqnarray*}
   U_{n}=O(n^{1-\delta}), \quad M_{n}\sqrt{\frac{\log n}{n}}w^{-1}(U_{n})=o(1), \quad n\to \infty
  \end{eqnarray*}
  for some positive number \( \delta,  \) where
\begin{eqnarray*}
    M_{n}&=& \sup_{\|v\|\leq U_{n}}|\phi_{\pi}^{-1}(v)|.
\end{eqnarray*}
\end{description}

\begin{thm}
\label{UpperBounds2}
 Suppose that the assumptions (AX), (AN), (AP), (AK1), (AK2) and (AH) hold.
Let \( \widetilde\rho^{(0)}_{n} \) be the estimate for the transformed L\'evy density \(\widetilde\rho^{(0)}\) defined in Section~\ref{rhotilde} and let  $w$ be a monotone positive Lipschitz function on $\mathbb{R}_{+}$ satisfying
\begin{equation}
\label{decreasing_w_thm}
0<w(z)\leq \log^{-1}(1+z), \quad z\in \mathbb{R}_{+}.
\end{equation}
If \( \widetilde\nu^{(0)}\in \widetilde{\mathfrak{L}}_\varkappa \)
for some \( 1\leq \varkappa<2, \)  then
\begin{eqnarray*}
     \left\|\widetilde\rho^{(0)}-\widetilde\rho^{(0)}_{n}\right\|_{L_{\infty}(\mathbb{R}^{m})}=O_{a.s.}\left( \sqrt{\frac{\log  n}{n}}\int_{[-U_{n},U_{n}]^{m}}w^{-1}(\| v \|) \mathfrak{R}(v)\,dv+U_{n}^{-(\varkappa-1)}\right),
\end{eqnarray*}
where \( \mathfrak{R}(v)=|\phi_{\pi}(v)|^{-1}\int_{[-1,1]^{m}}|1-\phi_{\pi}(v)/\phi_{\pi}(v+w)|\, dw. \)
\end{thm}
\begin{cor}
Consider a class of affine models \( \mathcal{A}(\bar \sigma,\varkappa ), \) such that  \( \nu^{(0)}\in \widetilde{\mathfrak{L}}_{\varkappa} \) and
\begin{eqnarray}
\label{phi_exp}
    |\phi_{\pi}(v)|\geq A\exp(-\bar\sigma^{2}\| v \|_{2}^{2}), \quad v\in \mathbb{R}^{m}
\end{eqnarray}
with some constant \( A>0. \) Then under the choice
\[
U_{n}=m^{-1}\bar\sigma^{-1}\sqrt{\frac{1}{2}\log n+((m+1)/2+\varkappa-1)\log \log n },
\]
it holds
 \begin{eqnarray*}
     \left\|\widetilde\rho^{(0)}-\widetilde\rho^{(0)}_{n}\right\|_{L_{\infty}(\mathbb{R}^{m})}=O_{a.s.}\left( \log^{-(\varkappa-1 )/2} n\right).
\end{eqnarray*}
\end{cor}
\begin{rem}
The condition \eqref{phi_exp} holds if, for example,  \( \lambda_{\max}(\mathfrak{A})\leq \bar\sigma^{2} \) with \(\mathfrak{A}:= (\alpha^{(0)}_{ij})_{m+1\leq i,j \leq d}. \)
\end{rem}
\subsection{Lower risk bounds}
The rates of Theorem~\ref{UpperBounds2} can not be improved in general as the following theorem states
\begin{prop}
\label{lower_bounds}
The following minimax lower risk bounds hold
\begin{eqnarray*}
    \liminf_{n\to \infty}\inf_{\widetilde\rho^{(0)}_{n}}\sup_{\chi\in\mathcal{A}(\bar \sigma, \varkappa)}\P_{\chi}\left( (\log n)^{(\varkappa-1)/2}\sup_{x\in \mathcal{D}}|\widetilde\rho^{(0)}(x)-\widetilde\rho^{(0)}_{n}(x)|  >\epsilon \right)>0,
\end{eqnarray*}
 where \( \varepsilon\) is any positive number, \(\widetilde\rho^{(0)}_{n} \) is any estimator of \( \widetilde\rho^{(0)}\)  based on \( n \) observations and the supremum is taken over all affine models \( \chi \) from the class \( \mathcal{A}(\bar \sigma, \varkappa). \)
\end{prop}
\section{Numerical example}
\label{NE}
\begin{figure}[pth]
\centering \includegraphics[width=12cm]{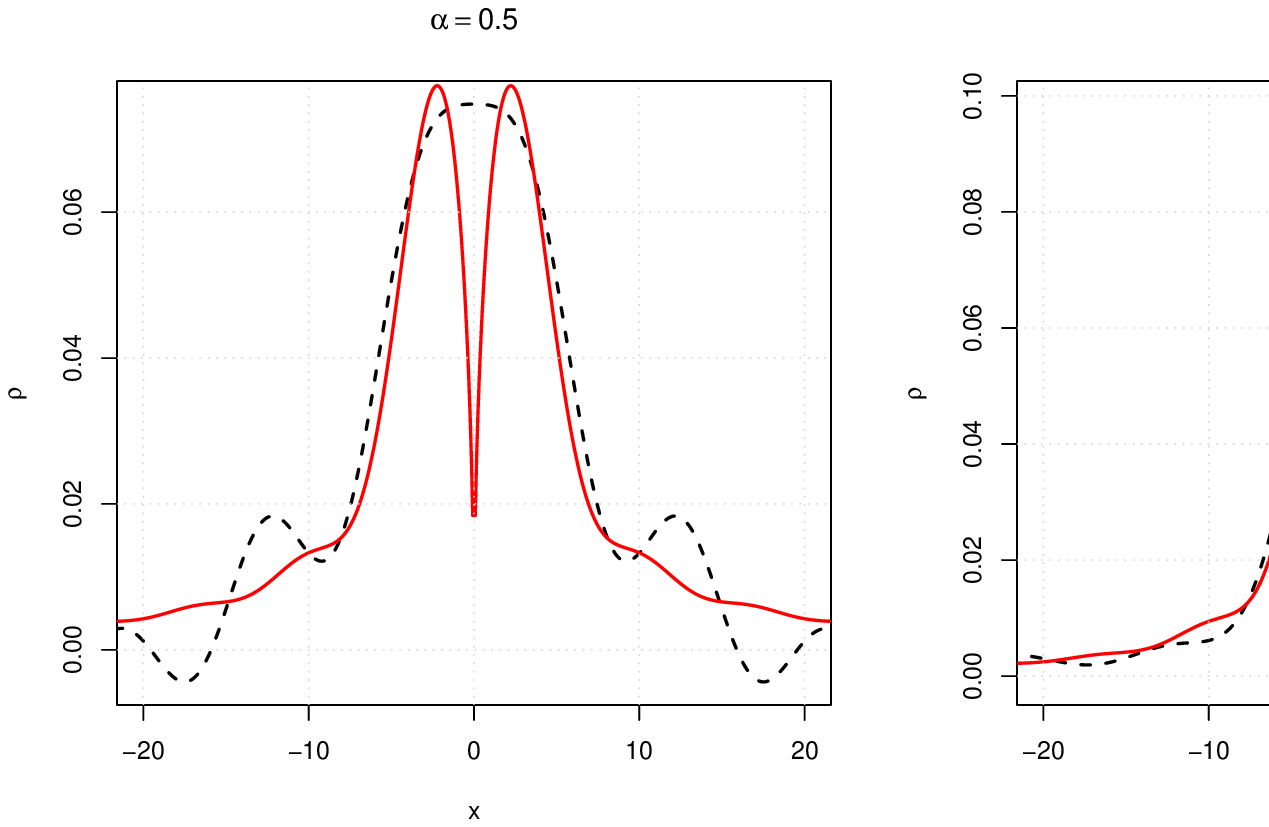}
\caption{ Typical estimates for  transformed L\'evy density \( \rho  \) (dashed black line) together with the true \( \rho  \) (solid red line) in the Bates stochastic volatility model with symmetric stable process \( Z_{t} \) and different stability indexes \( \alpha  \). }%
\label{levy_dens_est}%
\end{figure}

\begin{figure}[pth]
\centering \includegraphics[width=12cm]{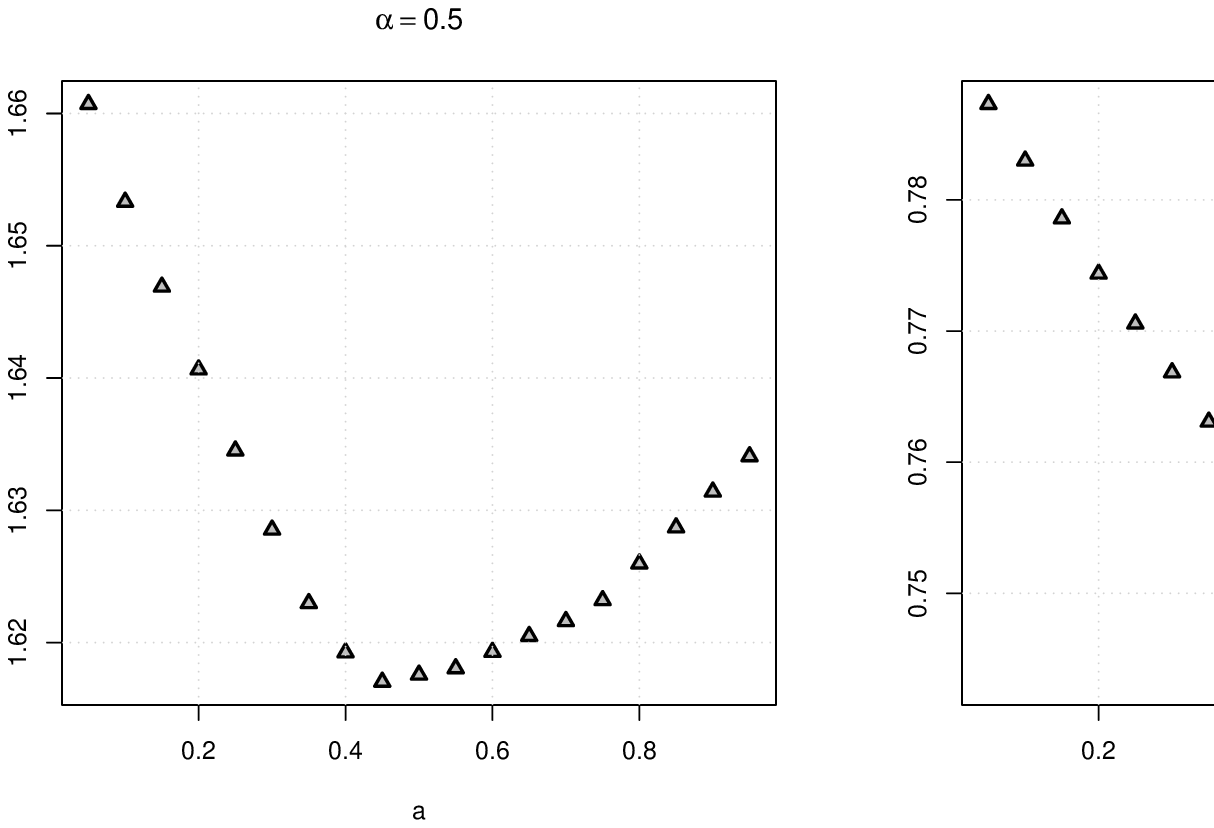}
\caption{ Function \( \mathcal{O}(a)  \)  in the case of symmetric stable process \( Z_{t} \) with stability
indexes \( 0.5 \) (left) and \( 0.8 \) (right) respectively.  }%
\label{alpha_objective}%
\end{figure}

\begin{figure}[pth]
\centering \includegraphics[width=12cm]{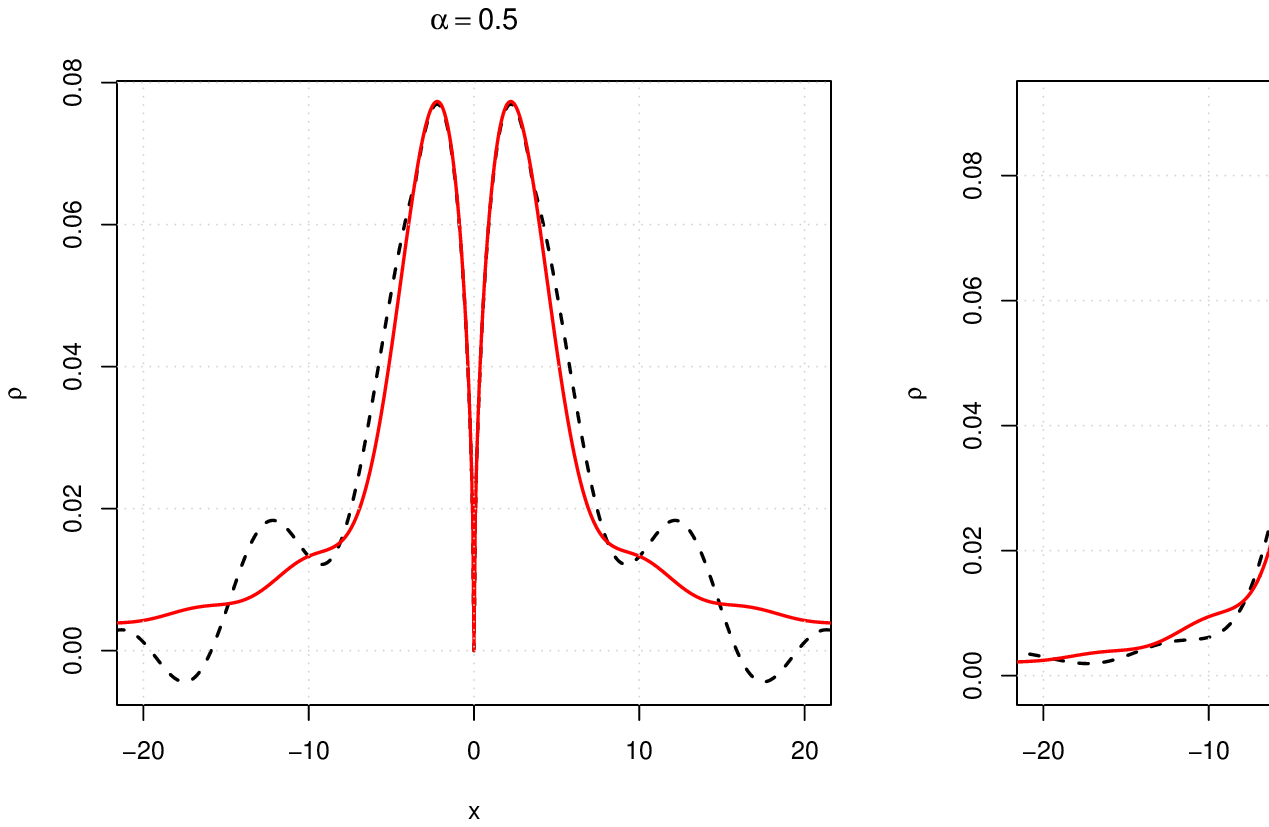}
\caption{ Corrected estimates for transformed L\'evy density \( \rho  \) (dashed black line) together with the true \( \rho  \) (solid red line) in the Bates stochastic volatility model with symmetric stable process \( Z_{t} \) and different stability indexes \( \alpha  \). }%
\label{levy_dens_est_corr}%
\end{figure}

Let us consider  a class of stochastic volatility models of the type
\begin{eqnarray}
    \label{Bates}
    dX(t)&=&-\frac{1}{2}V(t)\,dt+\sqrt{V(t)}dW^{S}(t)+dZ_{t},
    \\
    \nonumber
    dV(t)&=&\lambda (\theta-V(t))dt+\zeta\sqrt{V(t)}dW^{V}(t),
\end{eqnarray}
where  \( W^{S} \) and \( W^{V} \) are two independent Brownian motions, \( \lambda, \theta, \zeta \) are positive constants and \( Z_{t} \) is a pure-jump L\'evy process
with L\'evy density \( \nu(x).  \) This is a special type of the  model introduced in \citet{B1} that satisfies our assumptions.
In our numerical example we take \( Z(t) \) to be \( \alpha  \)-stable L\'evy process with stability index \( \alpha<1  \), i.e.,
\begin{eqnarray*}
    \nu(x)=C/|x|^{1+\alpha }, \quad \rho(x)=2\left( 1- \frac{\sin x}{x} \right) \nu(x)
\end{eqnarray*}
for some constant \( C>0. \)
For the sake of simplicity we consider a fixed design and  simulate a set of i.i.d pairs
\begin{equation}
\label{BSample}
 (X^{(k)}(0),X^{(k)}(\Delta)),
 \quad k=1,\ldots,n,
 \end{equation}
 with some fixed \( \Delta>0 \), where
 \[
 X^{(1)}(0)=\ldots=X^{(n)}(0)=0.
 \]
 Our aim is to reconstruct
 \( \rho  \) using the sample \eqref{BSample}. First, compute
\begin{eqnarray*}
\psi_{s,n}(v)&:=&\frac{1}{n\Delta} \sum_{j=1}^{n}\left[ \exp\left(\i vX^{(j)}(\Delta)\right)-\exp\left(\i v X^{(j)}(0)\right) \right]
\end{eqnarray*}
and
\begin{eqnarray*}
  \Psi_{0,n}(v)&:=&\int_{-1}^{1} \left(\psi_{s,n}(v)- \psi_{s,n}(v+w)\right) \, dw.
\end{eqnarray*}
\begin{rem}
In the case of mixed-frequency data observations are usually available for different frequency scales \( \Delta \) and
the choice of an appropriate frequency for estimation procedure  should be done depending on \( n \),
the number of points available for the given frequency scale. If \( \Delta \) is too
small then the variance of   \( \phi_{s,n}(v) \) explodes.
On the other hand, if \( \Delta \)   is too large than the approximation error of \( \phi_{s,n}(v) \) becomes large.
\end{rem}
Next define a parametric family of functions
\begin{eqnarray*}
 \rho^{(0)}_{n}(x;U)&=&\Re\left\{ \frac{1}{2\pi}\int_{-U}^{U}e^{-\i ux}\left[ \Psi_{0,n}(u)- \Psi_{0,n}(U)  \right]\, du \right\}, \quad U>0
\end{eqnarray*}
and find \( U \) by solving the following minimization problem
\begin{eqnarray*}
 \widehat U=\arginf_{U}  \left\{  \int_{\{ |u|>U \}} \left|  \Psi_{0,n}(u)-  \Psi_{0,n}(U)  \right|^{2}\, du+\pi \int |\partial_{xx}\rho^{(0)}_{n}(x;U)|\,dx \right\},
\end{eqnarray*}
where \( \pi>0  \) is a regularization parameter. In fact, this approach for choosing \( U \)  employs additional information about
 smoothness of \( \rho \) and turns out to be rather efficient in practice.
 In Figure~\ref{levy_dens_est} typical results of estimation based on \(n=1000 \) samples \eqref{BSample}
 with \( \Delta=0.1 \) are shown for two specifications of the process \( Z_{t}. \)
As can be seen the overall quality of estimation is good taking into account severely ill-posedness of the underlying
 estimation problem. However, the behavior of the transformed L\'evy density \( \rho  \) at zero has not been captured by the estimation method. In order to correct \( \rho(x)  \) at \( x=0, \) we separately estimate the stability index \( \alpha  \) using
a modification of the spectral algorithm proposed in \citet{Bel} for L\'evy processes.
Motivated by relations \eqref{FF0} and \eqref{FF1}, we define for any \( a \in (0,1) \)
\begin{eqnarray}
\label{lin_regr}
\mathcal{O}(a ):=\min_{(l_{0},l_{1},l_{2},l_{3})}\int_{0}^{\widehat U}(\psi_{s,n}(u|0)-l_{3}u^{a }-l_{2}u^{2}-l_{1}u-l_{0})^{2}\,du
\end{eqnarray}
and estimate \( \alpha  \) via \( \widetilde \alpha :=\argmin_{a\in (0,1) } \mathcal{O}(a ). \)
In Figure~\ref{alpha_objective} functions \(  \mathcal{O}(a ) \) based on the same samples as in Figure~\ref{levy_dens_est} are shown.
The resulting estimates for \( \alpha  \) are \( \widetilde \alpha=0.451 \) and \( \widetilde \alpha=0.783 \)
respectively. Now we correct the estimate \(\rho^{(0)}_{n}(x;\widehat U)  \)  by setting
\begin{eqnarray*}
  \widehat\rho^{(0)}_{n}(x;\widehat U)=
  \begin{cases}
  c(\varepsilon)\left(1-\frac{\sin x}{x}\right)|x|^{-(1+\widetilde\alpha)}, & |x|\leq \epsilon,
  \\
  \rho^{(0)}_{n}(x;\widehat U), &  |x|>\epsilon,
  \end{cases}
\end{eqnarray*}
where for any \( \epsilon>0 \) the constant \( c(\epsilon) \) is chosen in such a way that
function \(\widehat\rho^{(0)}_{n}(x;\widehat U) \)   is continuous. Finally, we find small enough \( \epsilon>0 \) which minimizes the integral \( \int |\partial_{xx}\widehat\rho^{(0)}_{n}(x;\widehat U)|\,dx \). Here again the smoothness of
\( \rho \) is used. A corrected estimate  \(\widehat\rho^{(0)}_{n}(x;\widehat U) \) is shown in Figure~\ref{levy_dens_est_corr}.
\section{Exponential inequalities for dependent sequences}
\label{ExpIneq}
The following theorem can be  found in \citet{MPR}.
\begin{thm}
\label{EIB}
Let \( (Z_k, \, k\geq 1) \) be a strongly mixing sequence of centered real-valued random variables on the probability space $(\Omega,\cal F,P)$
with the mixing coefficients satisfying
\begin{eqnarray}
\label{ALPHA_EXP_DECAY}
   \alpha(n)\leq \bar\alpha\exp(-cn ),\quad n\geq 1,\quad \bar\alpha>0,\quad c>0.
\end{eqnarray}
Assume that $\sup_{k\geq 1}|Z_k|\leq M$ a.s.,
then there is a positive constant \( C \) depending on \( c \) and \( \bar \alpha \) such that
$$
\P\left\{ \sum_{i=1}^n Z_i\geq \zeta  \right\}\leq \exp\left[-\frac{C\zeta^2 }{nv^{2}+M^{2} +M\zeta \log^{2}(n)}\right].
$$
for all \( \zeta>0 \) and \( n\geq 4, \)
where
\begin{eqnarray*}
    v^{2}=\sup_{i}\left( \E[Z_{i}]^{2}+2\sum_{j\geq i}\Cov(Z_{i},Z_{j}) \right).
\end{eqnarray*}
\end{thm}
\begin{cor}
\label{COV_EST}
Denote
\[
\rho _{j}=\E\left[ Z_{j}^{2}\log ^{2(1+\varepsilon )}\left( |Z_{j}|^{2}\right) \right], \quad j=1,2,\ldots,
\]%
with arbitrary small \( \varepsilon>0 \) and suppose that all \( \rho_{j}  \) are finite. Then
\begin{eqnarray*}
    \sum_{j\geq i}\Cov(Z_{i},Z_{j})\leq C\max_{j}\rho _{j}
\end{eqnarray*}
for some constant \( C>0, \) provided  \eqref{ALPHA_EXP_DECAY} holds.
Consequently the following inequality holds
\begin{eqnarray*}
    v^{2}\leq \sup_{i}\E[Z_{i}]^{2}+C\max_{j}\rho _{j}.
\end{eqnarray*}
\end{cor}

\begin{proof}
Due to the Rio inequality%
\[
\left\vert \Cov(Z_{i},Z_{j})\right\vert \leq 2\int_{0}^{\alpha
(|j-i|)}Q_{Z_{i}}(u)Q_{Z_{j}}(u)du
\]%
where for any random variable $X$ we denote by $Q_{X}$ the quantile function
of $X.$ Define

\[
\rho _{X}=\E\left[ X^{2}\log ^{2(1+\varepsilon )}\left( |X|^{2}\right) \right].
\]%
 The Markov inequality implies for small enough $u>0$

\begin{eqnarray*}
\P\left( |X|>\frac{\rho _{X}^{1/2}}{u^{1/2}|\log (u)|^{(1+\varepsilon )}}%
\right)  &\leq &\E\left[ X^{2}\log ^{2(1+\varepsilon )}\left( |X|^{2}\right) )%
\right] \frac{\rho _{X}^{-1}}{u^{-1}\log ^{-2(1+\varepsilon )}(u)}
\\
&&\times
\log
^{-2(1+\varepsilon )}\left( \frac{\rho _{X}}{u\log ^{2(1+\varepsilon )}(u)}%
\right)  \\
&=&u\log ^{-2(1+\varepsilon )}\left( \rho _{X}\log ^{-2(1+\varepsilon
)}(u)\right) \leq u
\end{eqnarray*}%
and therefore
\[
Q_{X}(u)\leq \frac{\rho _{X}^{1/2}}{u^{1/2}|\log (u)|^{(1+\varepsilon )}}.
\]%
Hence%
\[
\left\vert \Cov(Z_{i},Z_{j})\right\vert \leq 2\int_{0}^{\alpha (|j-i|)}\frac{%
\sqrt{\rho _{i}\rho _{j}}}{u\log ^{2(1+\varepsilon )}(u)}du\leq 2%
\sqrt{\rho_{i}\rho _{j}}\log ^{-1-2\varepsilon }(\alpha (|j-i|))
\]%
and
\[
\sum_{j\geq i}\Cov(Z_{i},Z_{j})\leq C\sqrt{\rho_{i}\rho_{j}}%
\sum_{j>i}\frac{1}{|j-i|^{1+2\varepsilon }}
\]
with some constant \( C>0 \) depending on \( \bar\alpha.  \)
\end{proof}
\subsection{Bounds on large deviations probabilities for weighted sup norms}
\label{EXP_BOUNDS}

Let \( Z_{j},\)  \(j=1,\ldots, n, \) be a sequence of   random vectors in \( \mathbb{R}^{d} \) and let \( G_{n}(u,z), \) \( n=1,2,\ldots, \) be a sequence of complex-valued functions defined on \( \mathbb{R}^{d}\times \mathbb{R}^{d} \) for some natural number \( d. \)   Define
\begin{eqnarray*}
m_{n}(u)&=&\frac{1}{n}\sum_{j=1}^{n}G_{n}(u,Z_{j}).
\end{eqnarray*}
\begin{prop}
\label{ExpBounds}
Suppose that  the following assumptions hold:
\begin{description}
  \item[(AZ1)]  The sequence \( Z_{j}, \) \( j=1,\ldots, n, \) is strictly stationary and  is \( \alpha \)-mixing  with the mixing coefficients \( (\alpha_{Z}(k))_{k\in \mathbb{N}} \) satisfying
  \begin{eqnarray*}
    \alpha_{Z}(k)\leq \bar\alpha_{0}\exp(-\bar\alpha_{1} k),\quad k\in \mathbb{N},
  \end{eqnarray*}
  for some \( \bar\alpha_{0}>0 \) and \( \bar\alpha_{1}>0. \)
  \item[(AZ2)] It holds \( \E\| Z_{j} \|^{p}<\infty \) for some \( p>2. \)
  \item[(AG1)] Each function \( G_{n}(u,z), \, n\in \mathbb{N}, \) is  Lipschitz in \( u \)
  with (at most)  linearly growing (in \( z \))  Lipschitz constant,
   i.e., for any \( u_{1},u_{2} \in \mathbb{R}^{d} \)
  \begin{eqnarray*}
  |G_{n}(u_{1},z)-G_{n}(u_{2},z)|\leq L_{n}(a+b\|z\|)\|u_{1}-u_{2}\|
  \end{eqnarray*}
where \( a,b \) are two non-negative real numbers
not depending on \( n \) and the sequence \( L_{n} \) does not depend on \( u. \)
  \item[(AG2)] There are two sequences \( \mu_{n}  \) and \( \sigma_{n},  \)
  such that
\begin{eqnarray*}
    \E[|G_{n}(u,Z)|^{2}]\leq \sigma^{2}_{n}, \quad |G_{n}(u,z)|&\leq & \mu_{n}, \quad (u,z)\in \mathbb{R}^{2d}.
\end{eqnarray*}
and the following asymptotic relations are fulfilled
\begin{eqnarray*}
 \mu_{n}/\sigma^{2}_{n}=O(1), \quad  \mu_{n}/\sigma_{n}=O\left(n^{1/2}\log^{-4}n\right), \quad \sigma^{2}_{n}=O(n),  \quad  L_{n}=O(\mu_{n} ),
\end{eqnarray*}
as \( \quad n\to \infty. \)
\end{description}
 Let $w$ be  Lipschitz continuous, positive, monotone decreasing  on $\mathbb{R}_{+}$  function  such that
\begin{equation}
\label{w}
0<w(z)\leq \log^{-1/2}(e+z), \quad z\in \mathbb{R}_{+}.
\end{equation}
 Then  there is \( \delta'>0 \) and \( \xi_{0}>0  \), such that
the inequality
\begin{eqnarray}
\label{MINEQ}
\P\left\{\log^{-(1+\varepsilon)}(1+\mu_{n})\sqrt{\frac{n}{\sigma^{2}_{n}\log n}}
\sup_{u\in \mathbb{R}^{d}} \left[ w(\| u \|)| m_{n}(u)- \E[m_{n}(u)] | \right]>\xi \right\}&\leq & B n^{-1-\delta' }
\end{eqnarray}
holds for any \( \xi >\xi_{0},  \) some positive constant \( B \) depending on \( \xi  \) and arbitrary small \( \varepsilon>0. \)
\end{prop}
\begin{proof}
 Define for \( j=1,\ldots,n, \)
\begin{eqnarray*}
\zeta_{j}(u)&=&G_{n}(u,Z_{j})-\E[G_{n}(u,Z_{j})]
\end{eqnarray*}
and introduce the process
\begin{eqnarray*}
\mathcal{W}_{n}(u)=\frac{1}{n}\sum_{j=1}^{n}
w(\| u \|)\zeta_{j}(u).
\end{eqnarray*}

Consider the sequence \( A_{k}=e^{k},\, k\in \mathbb{N} \) and cover each
cube \( [-A_{k},A_{k}]^{d} \) with
\( M_{k}=\left(\lfloor (2d^{1/2}A_{k})/\gamma \rfloor +1  \right)^{d} \) disjoint small
cubes \( \Lambda_{k,1},\ldots,\Lambda_{k,M_{k}} \), the length of each cube being equal to \( \gamma/d^{1/2}. \) Let \( u_{k,1},\ldots, u_{k,M_{k}} \) be the centers
of these cubes. We have for any natural \( K>0 \)
\begin{eqnarray*}
\max_{k=1,\ldots,K}\sup_{A_{k-1}<\| u \|\leq A_{k}}|\mathcal{W}_{n}(u)|&\leq &
\max_{k=1,\ldots,K}\max_{\| u_{k,m} \|>A_{k-1}}|\mathcal{W}_{n}(u_{k,m})|
\\
&&+\max_{k=1,\ldots,K}\max_{1\leq m \leq M_{k}}
\sup_{u\in \Lambda_{k,m}}|\mathcal{W}_{n}(u)-\mathcal{W}_{n}(u_{k,m})|.
\end{eqnarray*}
Hence
\begin{multline}
\label{DEC1}
\P\left( \max_{k=1,\ldots,K}\sup_{A_{k-1}< \| u \|\leq A_{k}}|\mathcal{W}_{n}(u)|>\lambda \right)
\leq \sum_{k=1}^{K}\sum_{\{ \| u_{k,m} \|>A_{k-1} \}}\P(|\mathcal{W}_{n}(u_{k,m})|>\lambda/2)+
\\
\P\left(\sup_{\| u-v
\|<\gamma}|\mathcal{W}_{n}(v)-\mathcal{W}_{n}(u)|>\lambda/2\right).
\end{multline}
It holds for any \( u,v\in \mathbb{R}^{d} \)
\begin{eqnarray}
\label{WNDIFF}
\nonumber
|\mathcal{W}_{n}(v)-\mathcal{W}_{n}(u)|&\leq&
2\mu_{n}|w(\| v \|)-w(\| u \|)|
\\
\nonumber
&&+\frac{1}{n}\sum_{j=1}^{n}
\left| G_{n}(v,Z_{j})-G_{n}(u,Z_{j}) \right|
\\
\nonumber
&&+
\frac{1}{n}\sum_{j=1}^{n}
\left|\E \left[ G_{n}(v,Z_{j})-G_{n}(u,Z_{j}) \right] \right|
\\
\nonumber
&\leq& 2(L_{n}\vee \mu_{n}) \| u-v \|
\left[ L_{w}+\frac{1}{n}\sum_{j=1}^{n}(a+b\| Z_{j} \|)\right.
\\
&& \left.+\frac{1}{n}\sum_{j=1}^{n}(a+b\E \| Z_{j} \|) \right],
\end{eqnarray}
where \( L_{\omega } \) is the Lipschitz constant of \( w. \) By the Markov inequality
\begin{eqnarray*}
\P\left( \frac{1}{n}\sum_{j=1}^{n}[\| Z_{j} \|-\E \| Z_{j} \|]>
c \right)\leq c^{-p}n^{-p}\E\left| \sum_{j=1}^{n}[\| Z_{j} \|-\E \| Z_{j} \| ] \right|^{p}
\end{eqnarray*}
for any \( c >0. \)
Using the  moment  inequality of Yokoyama (see \citet{Y}),  we get
\begin{eqnarray*}
  \E\left| \sum_{j=1}^{n}[\| Z_{j}\|-\E\| Z_{j}\| ] \right|^{p}\leq C_{p}(\alpha)n^{p/2},
\end{eqnarray*}
where \( C_{p} (\alpha) \) is some constant depending on \( p \) and \( \alpha=(\bar\alpha_{0},\bar\alpha_{1}) \) from
the assumption (AZ1).
 Hence
\begin{eqnarray}
\label{MOMINEQ}
\P\left( \frac{1}{n}\sum_{j=1}^{n}\| Z_{j} \|>
2\beta_{1} \right)\leq  C_{p}(\alpha ) n^{-p/2}/\beta^{p}_{1}
\end{eqnarray}
with \( \beta_{1}=\E \| Z_{j} \|\).
 Setting
\begin{eqnarray*}
\gamma=\lambda/(4(L_{n}\vee \mu_{n})(2a+3b\beta_{1}+L_{\omega}))
\end{eqnarray*}
and combining \eqref{WNDIFF} with the inequality \eqref{MOMINEQ},
we obtain
\begin{eqnarray}
\label{LINEQ}
\P\left(\sup_{\| u-v
\|<\gamma}|\mathcal{W}_{n}(v)-\mathcal{W}_{n}(u)|>\lambda/2\right)\leq
B_{1}n^{-p/2}
\end{eqnarray}
with some constant \( B_{1} \)  depending neither on \( \lambda \) nor \( n \).
Turn now to the first term on the right-hand side of  \eqref{DEC1}.
If \( \| u_{k,m} \|>A_{k-1}, \) then it follows from Theorem~\ref{EIB} and Corollary~\ref{COV_EST}
\begin{multline*}
\P\left(\left|\Re\left[\mathcal{W}_{n}(u_{k,m}) \right]\right|>\lambda/4\right)
\\
\leq   B_{2}\exp\left(
-\frac{\lambda^{2}n}{B_{3} w^{2}(A_{k-1})\sigma^{2}_n \log^{2(1+\epsilon)}(1+\mu_{n}) +w^{2}(A_{k-1})\mu^{2} _{n}/n+\lambda\mu _{n}\log^{2}(n)
w(A_{k-1})}\right),
\end{multline*}
\begin{multline*}
\P\left(\left|\Im\left[\mathcal{W}_{n}(u_{k,m}) \right]\right|>\lambda/4\right)
\\
\leq  B_{4}\exp\left(
-\frac{\lambda^{2}n}{B_{3} w^{2}(A_{k-1})\sigma^{2}_n \log^{2(1+\epsilon)}(1+\mu_{n}) +w^{2}(A_{k-1})\mu^{2} _{n}/n+\lambda\mu _{n}\log^{2}(n)
w(A_{k-1})}\right)
\end{multline*}
with some constants \( B_{2} \), \( B_{3} \) and \( B_{4} \) depending only on the characteristics
of the process \( Z, \) and \( \varepsilon>0. \) Indeed, due to  (AG2) it holds
\begin{eqnarray*}
    \E\left[ |\zeta_{j}|^{2}\log^{2(1+\epsilon)}\left(|\zeta_{j}|^{2}\right)\mathbf{1}_{\{ |\zeta_{j}|>1 \}}\right]&\leq &  \E\left[ |G_{n}(u,Z_{j})|^{2}\right] \log^{2(1+\epsilon)}(2\mu_{n})
    \\
    &\leq & B_{5}\sigma_{n}^{2}  \log^{2(1+\epsilon)}(1+\mu_{n})
\end{eqnarray*}
with some constant \( B_{5}>0. \)
Hence
\begin{eqnarray*}
 \E\left[ |w\zeta_{j}|^{2}\log^{2(1+\epsilon)}\left(|w\zeta_{j}|^{2}\right) \right]&=&
 w^{2}\E\left[ |\zeta_{j}|^{2}\log^{2(1+\epsilon)}\left(|\zeta_{j}|^{2}\right) \right]
 +w^{2}\log^{2(1+\epsilon)}\left(w^{2}\right) \E[|\zeta_{j}|^{2}]
 \\
 &\leq &  w^{2}\left(1+B_{5}\sigma_{n}^{2} \log^{2(1+\epsilon)}(1+\mu_{n})\right)
 + w^{2}\log^{2(1+\epsilon)}\left(w^{2}\right)\sigma_{n}^{2}.
\end{eqnarray*}
Taking \( \lambda=\xi  n^{-1/2}\sigma_{n}\log^{1/2} (n) \log^{1+\epsilon} (1+\mu_{n})\) with
 some \( \xi>0 \), we get
\begin{multline*}
\sum_{\{ \| u_{k,m} \|>A_{k-1} \}}\P(|\mathcal{W}_{n}(u_{k,m})|>\lambda/2)\leq
\left(\lfloor (2d^{1/2}A_{k})/\gamma \rfloor +1  \right)^{d}
\\
\times\exp\left(
-\frac{\lambda^{2}n}{B_{3}w^{2}(A_{k-1})\sigma_{n}^{2} \log^{2(1+\varepsilon) }(1+\mu_{n} )+w^{2}(A_{k-1})\mu^{2} _{n}/n+\lambda\mu _{n}\log^{2}(n)
w(A_{k-1})}\right)
\\
\lesssim   A^{d}_{k} n^{d/2}\sigma^{-1}_{n}(L_{n}\vee \mu_{n})\log^{-3/2-\epsilon} (n)
\exp\left(-\frac{B\xi^{2}\log (n)}{w^{2}(A_{k-1})}\right), \quad n\to \infty
\end{multline*}
with some constant \( B>0. \)
Fix \( \theta>0 \) such that \( B\theta>1 \)  and compute
\begin{eqnarray*}
 \sum_{\{ \| u_{k,m} \|>A_{k-1} \}}\P(|\mathcal{W}_{n}(u_{k,m})|>\lambda/2)
& \lesssim &
e^{dk-\theta B(k-1)}n^{(d+1)/2} \log^{-3/2-\epsilon} (n) e^{-B(k-1)(\xi^{2}\log n-\theta)}
 \\
 &\lesssim&
 e^{k(1-\theta B)} \log^{-3/2-\epsilon} (n) e^{-B(k-1)(\xi^{2}\log n-\theta)+(d+1)\log(n)/2}.
\end{eqnarray*}
If \( \xi^{2}\log n>\theta \) we derive
\begin{eqnarray*}
 \sum_{k=2}^{K}\sum_{\{ \| u_{k,m} \|>A_{k-1} \}}\P(|\mathcal{W}_{n}(u_{k,m})|>\lambda/2)
& \lesssim &    \log^{-3/2-\epsilon} (n) e^{-(B\xi^{2}-(d+1)/2)\log(n)}.
\end{eqnarray*}
Taking \( \xi >\xi_{0}  \) for large enough \( \xi_{0} \), we get  \eqref{MINEQ}.

\end{proof}

\section{Proofs}
\label{proofs}
\subsection{Proof of Theorem~\ref{phi_conv}}
The smallest eigenvalue \( \lambda_{\Gamma} \) of the matrix \( \Gamma \) satisfies
\begin{eqnarray}
\label{lamdag}
\nonumber
\lambda_{\Gamma}&=&\min_{\| W \|=1}W^{\top}\Gamma W
\\
\nonumber
&\geq& \min_{\| W \|=1}W^{\top}\E[\Gamma] W+\min_{\| W \|=1}W^{\top}(\Gamma-\E [\Gamma]) W
\\
&\geq& \min_{\| W \|=1}W^{\top}\E[\Gamma] W-\sum_{k=0}^{2}\sum_{|\mathbf{m}_{1}|,|\mathbf{m}_{2}|\leq
l}|\Gamma_{k,\mathbf{m}_{1},\mathbf{m}_{2}}-\E[\Gamma_{k,\mathbf{m}_{1},\mathbf{m}_{2}}]|.
\end{eqnarray}
By the assumption (AG)
\begin{eqnarray*}
\min_{\| W \|=1}\left[ W^{\top}\E[\Gamma] W \right]\geq\gamma_{0}
\end{eqnarray*}
with some \( \gamma_{0}>0. \)
For \( j=1,\ldots, n, \) any \( k=0,1,2 \) and any multi-indices \( \mathbf{m}_{1} \), \( \mathbf{m}_{2} \)
such that \( |\mathbf{m}_{1}|, |\mathbf{m}_{2}|\leq l \), define
\begin{multline*}
\Delta_{j}=\frac{1}{h_{1}h_{2}^{m}}\left( \frac{\delta_{j}}{h_{1}} \right)^{k}\left( \frac{X^{(m)}(t_{j})-x}{h_{2}} \right)^{\mathbf{m}_{1}+\mathbf{m}_{2}}
K\left( \frac{\delta_{j}}{h_{1}}, \frac{X^{(m)}(t_{j})-x}{h_{2}} \right)
\\
-\int_{\mathbb{R}^{m}}\int_{\mathbb{R}_{+}}t^{k}z^{\mathbf{m}_{1}+\mathbf{m}_{2}}
K(t,z)p_{m,\pi}(x+h_{2}z)p_{\delta}(h_{1}t)\,dt\, dz.
\end{multline*}
We have \( \E [\Delta_{j}]=0 \),
\[
|\Delta_{j}|\leq h_{1}h_{2}^{-m}\sup_{t\in \mathbb{R}_{+}}\sup_{z\in \mathbb{R}^{m}}\left[(1+t^{2}) (1+\| z \|^{2l})K(t,z)
\right]=:K_{1}h_{1}h_{2}^{-m}
\]
and
\begin{eqnarray*}
\E[\Delta_{j}]^{2}&\leq & \int_{\mathbb{R}^{m}}\int_{\mathbb{R}_{+}}t^{2k}z^{2u_{1}+2u_{2}}
K^{2}(t,z)p_{m,\pi}(x+h_{2}z)p_{\delta}(h_{1}t)\,dt\, dz
\\
&\leq & \frac{\bar p_{m,\pi} \bar p_{\delta}}{h_{1}h_{2}^{m}}\int_{\mathbb{R}^{m}}(1+t^{4})(1+\| z \|^{4l})K^{2}(t,z)\,dz=:
K_{2}h_{1}^{-1}h_{2}^{-m},
\end{eqnarray*}
where \( \bar p_{X}=\sup_{z\in \mathbb{R}^{m}} p_{m,\pi}(z), \) \( \bar p_{\delta}=\sup_{s\in [0,T]}p_{\delta}(s) \)
and \( K_{1}, K_{2} \) are two positive constants.
According to Proposition~\ref{ExpBounds}, we have for any \( \zeta>0 \)
\begin{eqnarray}
\label{EIGamma}
\P\left( |\Gamma_{k,\mathbf{m}_{1},\mathbf{m}_{2}}-\E[\Gamma_{k,\mathbf{m}_{1},\mathbf{m}_{2}}]|
\geq \zeta  \right)
=\P \left( \frac{1}{n}
\left| \sum_{j=1}^{n}\Delta_{j} \right|\geq\zeta \right)
\leq D_{0}\exp(-\zeta B_{0} n h_{1}h_{2}^{m})
\end{eqnarray}
with some positive constants \( D_{0} \) and \( B_{0} \).
Combining  \eqref{lamdag} with \eqref{EIGamma}, we get
\begin{eqnarray*}
    \P\left(\lambda_{\Gamma}
    \leq\gamma_{0}/2\right)\leq  D_{0}N^{2}_{l}\exp(-\gamma_{0} B_{0} n
    h_{1}h_{2}^{m}/2N^{2}_{l}),
\end{eqnarray*}
where \( N^{2}_{l} \) is the number of elements in the matrix \( \Gamma. \)
Introduce the matrices \( Q_{k}=(Q_{k,j,\mathbf{m}})_{1\leq j \leq n,\, |\mathbf{m}|\leq l} \) with elements
\begin{eqnarray*}
    Q_{k,j,\mathbf{m}}=\left( \frac{\delta_{j}}{h_{1}} \right)^{k}\left(\frac{X^{(m)}(t_{j})-x}{h_{2}}\right)^{\mathbf{m}}\sqrt{\frac{1}{nh_{1}h_{2}^{m}}K\left(\frac{\delta_{j} }{h_{1}}, \frac{X^{(m)}(t_{j})-x}{h_{2}} \right)}.
\end{eqnarray*}
Set
\begin{eqnarray*}
     \phi_{m}(v|s,x)=\phi\left(v,0\ldots,0|s,\E[X_{m+1}(0)],\ldots,\E[X_{d}(0)]\right).
\end{eqnarray*}
Denote by \( Q_{k,\mathbf{m}} \) the \( \mathbf{m} \)th column of \( Q_{k}, \) \( k=0,1, \) and
define
\begin{eqnarray*}
    Q_{0}^{C}&=&\sum_{|\mathbf{m}|\leq l} \frac{\partial_{x^{\mathbf{m}}}\phi_{m}(v|0,x)h_{2}^{\mathbf{m}}}{\mathbf{m}!}Q_{0,\mathbf{m}},
    \\
    Q_{1}^{C}&=&h_{1}\sum_{|\mathbf{m}|\leq l} \frac{\partial_{x^{\mathbf{m}}}\partial_{s}\phi_{m}(v|0,x)h_{2}^{\mathbf{m}}}{\mathbf{m}!}Q_{1,\mathbf{m}},
\end{eqnarray*}
\begin{eqnarray*}
     Q=
     \begin{pmatrix}
    Q_{0} \\
    Q_{1}
    \end{pmatrix}, \quad
    Q^{C}= \begin{pmatrix}
    Q^{C}_{0} \\
    Q^{C}_{1}
    \end{pmatrix}.
\end{eqnarray*}
Since \(\Gamma=Q^{\top}Q \), we get  \((P(0),P(0))^{\top}
\Gamma^{-1}Q^{\top}Q^{C}=(\phi_{m}(v|0,x),h_{1}\phi_{m}(v|0,x)) \).
Thus, we can write
\begin{eqnarray*}
    (\phi_{n}(v)-\phi_{m}(v|0,x),h_{1}(\phi_{s,n}(v)-\phi_{m,s}(v|0,x)))^{\top}
    &=&(P(0),P(0))^{\top}\Gamma^{-1}(S-Q^{\top}Q^{C})
    \\
    &=&(P(0),P(0))^{\top}\Gamma^{-1}\varepsilon_{n}(v),
\end{eqnarray*}
where \( \varepsilon_{n}(v)=(\varepsilon_{n,0}(v), \varepsilon_{n,1}(v)) \) and
\begin{eqnarray*}
    \varepsilon_{n,k,\mathbf{m}}(v)&=&\frac{1}{nh_{1}h_{2}^{m}}\sum_{j=1}^{n}\left[Z_{j}(v)
    -\bar{\mathcal{Q}}_{0}(X^{(m)}(t_{j})-x)-\delta_{j}\bar{\mathcal{Q}}_{1}(X^{(m)}(t_{j})-x)\right]\times
    \\
    && \times \left( \frac{\delta_{j}}{h_{1}} \right)^{k}\left( \frac{X^{(m)}(t_{j})-x}{h_{2}}\right)^{\mathbf{m}}K\left(\frac{\delta_{j}}{h_{1}}, \frac{X^{(m)}(t_{j})-x}{h_{2}} \right),
\end{eqnarray*}
where
\begin{eqnarray*}
 \bar{\mathcal{Q}}_{0}(z)=\sum_{|\mathbf{m}|\leq l} \frac{\partial_{x^{\mathbf{m}}}\phi_{m}(v|0,x)z^{\mathbf{m}}}{\mathbf{m}!}
\end{eqnarray*}
and
\begin{eqnarray*}
 \bar{\mathcal{Q}}_{1}(z)=\sum_{|\mathbf{m}|\leq l} \frac{\partial_{x^{\mathbf{m}}}\partial_{s}\phi_{m}(v|0,x)z^{\mathbf{m}}}{\mathbf{m}!}.
\end{eqnarray*}
So, on the set \( \{ \lambda_{\Gamma}>\gamma_{0}/2 \} \) we get
\begin{eqnarray*}
  \max\left\{ |\phi_{n}(v)-\phi_{m}(v|0,x)|,|\phi_{s,n}(v)-\phi_{m,s}(v|0,x)| \right\}&\leq & \| \Gamma \varepsilon_{n} \|\leq
  \lambda_{\Gamma}^{-1}\| \varepsilon_{n} \|
  \\
  &\leq& 2\gamma_{0}^{-1}\| \varepsilon_{n} \|
  \\
  &\leq&
  2\gamma_{0}^{-1}N^{1/2}_{l}\max_{k,\mathbf{m}}|\varepsilon_{n,k,\mathbf{m}}(v)|.
\end{eqnarray*}
Denote
\begin{eqnarray*}
    \Delta^{(1)}_{j,k,\mathbf{m}}(v)&:=&h^{-1}_{1}h_{2}^{-m}\left[Z_{j}(v)
    -\phi_{m}(v|\delta_{j},X^{(m)}(t_{j}))\right]\times
    \\
    && \times \left( \frac{\delta_{j}}{h_{1}} \right)^{k}\left( \frac{X^{(m)}(t_{j})-x}{h_{2}}\right)^{\mathbf{m}}K\left(\frac{\delta_{j}}{h_{1}}, \frac{X^{(m)}(t_{j})-x}{h_{2}} \right),
    \\
    \Delta^{(2)}_{j,k,\mathbf{m}}(v)&:=&h^{-1}_{1}h_{2}^{-m}\left[\phi_{m}(v|\delta_{j},X^{(m)}(t_{j}))
    -\bar{\mathcal{Q}}_{0}(X^{(m)}(t_{j})-x)-\delta_{j}\bar{\mathcal{Q}}_{1}(X^{(m)}(t_{j})-x)\right]\times
    \\
    && \times \left( \frac{\delta_{j}}{h_{1}} \right)^{k}\left( \frac{X^{(m)}(t_{j})-x}{h_{2}}\right)^{\mathbf{m}}K\left(\frac{\delta_{j}}{h_{1}}, \frac{X^{(m)}(t_{j})-x}{h_{2}} \right).
\end{eqnarray*}
It holds
\begin{eqnarray*}
    |\varepsilon_{n,k,\mathbf{m}}(v)|\leq \left| \frac{1}{n}\sum_{j=1}^{n}\Delta_{j,k,\mathbf{m}}^{(1)}(v) \right|+\left| \frac{1}{n}\sum_{j=1}^{n}\left[ \Delta_{j,k,\mathbf{m}}^{(2)}(v)-\E\Delta_{j,k,\mathbf{m}}^{(2)}(v) \right] \right|+|\E\Delta_{j,k,\mathbf{m}}^{(2)}(v)|.
\end{eqnarray*}
Note that \( \E  \left[ \Delta_{j,k,\mathbf{m}}^{(1)}(v) \right]=0\) and
\begin{eqnarray*}
    |\Delta_{j,k,\mathbf{m}}^{(1)}(v)|\leq A_{11}h_{1}^{-1}h_{2}^{-m},\quad \E \left[ \Delta_{j,k,\mathbf{m}}^{(1)}(v) \right]^{2}\leq A_{12}h_{1}^{-1}h_{2}^{-m},
    \\
    \left| \Delta_{j,k,\mathbf{m}}^{(2)}(v)-\E\left[ \Delta_{j,k,\mathbf{m}}^{(2)}(v) \right] \right|\leq A_{21}h_{2}^{l-m},
    \quad \E \left[ \Delta_{j,k,\mathbf{m}}^{(2)}(v) \right]^{2}\leq A_{22}h_{2}^{2l-m}
\end{eqnarray*}
with some positive constants \( A_{11} \), \( A_{12} \), \( A_{21} \) and \( A_{22} \) not depending
on \( x \).
Proposition~\ref{ExpBounds} implies that
\begin{eqnarray*}
    \P\left( \sup_{v\in \mathbb{R}^{m}}\left\{ w(\| v \|)\left| \frac{1}{n}\sum_{j=1}^{n}\Delta_{j,k,\mathbf{m}}^{(1)}(v) \right| \right\}>A_{1}\frac{\log^{3/2+\epsilon} n}{\sqrt{nh_{1}h_{2}^{m}}} \right)\leq B_{1} n^{-1-\delta' }
\end{eqnarray*}
and
\begin{eqnarray*}
   \P\left( \sup_{v\in \mathbb{R}^{m}}\left\{ w(\| v \|)\left| \frac{1}{n}\sum_{j=1}^{n}\left(\Delta_{j,k,\mathbf{m}}^{(2)}(v)-\E\left[ \Delta_{j,k,\mathbf{m}}^{(2)}(v)\right] \right)\right| \right\}>A_{2}\frac{\log^{3/2+\epsilon} n}{\sqrt{nh_{2}^{2l-m}}} \right)\leq B_{2} n^{-1-\delta' }
\end{eqnarray*}
for some positive constants \( A_{1}, \) \( B_{1}, \) \( A_{2} \) and \( B_{2} \) not depending on \( k \) and \( \mathbf{m}. \)
Due to Lemma~\ref{RegAffine} and Assumption (AN), the function \( \phi_{m}(v|s,x) \) is at least two times differentiable in \( s. \) Hence
\begin{multline*}
    \sup_{|t|\leq h_{1}}\sup_{\|z-x\|\leq h_{2}}|\phi_{m}(v|0,z)
    -\bar{\mathcal{Q}}_{0}(z-x)-t\bar{\mathcal{Q}}_{1}(z-x)|
    \\
    \leq C\left[ h_{1}^{2}\sup_{|t|\leq h_{1}}\sup_{\|z-x\|\leq h_{2}}|\partial_{ss}\phi_{m}(v|t,x)|+ h_{2}^{l+1}\max_{|\mathbf{m}|\leq l+1}\sup_{\|z-x\|\leq h_{2}}|\partial_{x^{\mathbf{m}}}\phi_{m}(v|0,x)|+
     \right.
    \\
    \left.
    +h_{1}h_{2}^{l+1}\max_{|\mathbf{m}|\leq l+1}\sup_{\|z-x\|\leq h_{2}}|\partial_{x^{\mathbf{m}}}\partial_{s}\phi_{m}(v|0,x)|\right]
\end{multline*}
for some constant \( C>0. \)
Using Lemma~\ref{RegAffine}, we get \( |\E\Delta_{j,k,\mathbf{m}}^{(2)}(v)|\leq A_{3}(h_{1}^{2}+h_{2}^{l+1})\| v \|_{2}^{2(l+2)}, \) \( \| v \|_{2}>1. \)
\subsection{Proof of Theorem~\ref{UpperBounds1}}
Theorem~\ref{phi_conv} implies
\begin{eqnarray}
\label{PsiIneq}
    \P\left( \sup_{v\in \mathbb{R}^{m}}\left[ w(\| v \|)|\Psi_{0}(v)-\Psi_{0,n}(v)| \right]> 2^{m+1}A\zeta_{n}\right)\leq B n^{-1-\delta}.
\end{eqnarray}
Using the identity \( \mathcal{L}=\Psi_{0}(v)-\mathcal{F}[\rho^{(0)}](v), \) we get
\begin{eqnarray*}
 \mathcal{L}_{n}-\mathcal{L}&=&\int_{\mathbb{R}^{m}}U_{n}^{-m}\mathcal{K}_{\mathcal{L}}(v/U_{n})\left( \Psi_{0,n}(v)- \Psi_{0}(v)\right)\, dv+\int_{\mathbb{R}^{m}}U_{n}^{-m}\mathcal{K}_{\mathcal{L}}(v/U_{n})\mathcal{F} [\rho^{(0)}] (v)\, dv
\end{eqnarray*}
and hence
\begin{equation*}
|\mathcal{L}_{n}-\mathcal{L}|\leq \int_{\mathbb{R}^{m}}U_{n}^{-m}|\mathcal{K}_{\mathcal{L}}(v/U_{n})|
\left|\Psi_{0,n}(v)- \Psi_{0}(v) \right|\, du
 +U_{n}^{-m}\int_{\mathbb{R}^{m}}\left| \mathcal{K}_{\mathcal{L}}(v/U_{n})\mathcal{F} [\rho^{(0)}] (v) \right|\, dv.
\end{equation*}
By the inequality \eqref{ass_nu}
\begin{multline}
\label{LOBias}
\int_{\mathbb{R}^{m}}\left|\mathcal{K}_{\mathcal{L}}(v/U_{n})\mathcal{F} [\rho^{(0)}] (v) \right|\, dv\leq
C\int_{\mathbb{R}^{m}} \left[ |v_{1}|^{-\varkappa}\times\ldots\times |v_{d}|^{-\varkappa} \right]|\mathcal{K}_{\mathcal{L}}(v/U_{n})| \, dv
\leq C_{1}U_{n}^{-m(\varkappa-1)}
\end{multline}
with some constant \( C_{1}>0 \).
Combining \eqref{LOBias} with \eqref{PsiIneq}, we get
\begin{eqnarray*}
  U_{n}^{m}|\mathcal{L}_{n}-\mathcal{L}|=O_{a.s}\left(\zeta_{n}\int_{[-U_{n},U_{n}]^{m}}w^{-1}(\| v \|)\,dv
  +U_{n}^{-m(\varkappa-1)}\right).
\end{eqnarray*}
Finally, using the Fourier inversion formula, we derive
\begin{eqnarray*}
 \sup_{z\in \mathbb{R}^{m}}|\rho^{(0)}(z)-\rho^{(0)}_{n}(z)| &\leq & \left[ \int_{\mathbb{R}^{m}}|\mathcal{K}_{\rho}(v/U_{n})|\left| \Psi_{0,n}(v)-\Psi_{0}(v) \right|\,dv +U_{n}^{m}|\mathcal{L}_{n}-\mathcal{L}| \right]
    \\
    && +\left| \int_{\{ \| v \|>a_{K}U_{n} \}}\mathcal{F} [\rho^{(0)}] (v)(1-\mathcal{K}_{\rho}(v/U_{n}))\, dv \right|.
\end{eqnarray*}
\subsection{Proof of Theorem~\ref{UpperBounds2}}

Fix some \( D >0 \) and consider the event
\begin{eqnarray*}
  \mathcal{A}&=&\left\{ \sup_{\|v\|\leq U_{n}}\left[ w(\|v\|)|\phi(v)-\phi_{n}(v)| \right]\leq D \sqrt{\frac{\log n}{n}}\right\}.
\end{eqnarray*}
By the assumption (AH), it holds on \( \mathcal{A} \)
 \begin{eqnarray*}
\sup_{\|v\|<U_{n}}\left| \frac{\phi(v)-%
\phi_{n}(v)}{\phi(v)} \right|\leq D M_{n}w^{-1}(U_{n})\sqrt{\frac{\log n}{n}}=o(1),\quad n\to \infty
\end{eqnarray*}
and hence
\begin{eqnarray*}
    \log \phi_{n}(v)-\log\phi(v) =\log \left[ 1+\frac{\phi_{n}(v)-\phi(v)}{\phi(v)} \right] =\frac{\phi_{n}(v)-\phi(v)}{\phi(v)} +\mathcal{R}(v)
\end{eqnarray*}
with
\begin{eqnarray*}
|\mathcal{R}(v)|\leq  c\left| \frac{\phi_{n}(v)-\phi(v)}{\phi(v)} \right|^{2}
\end{eqnarray*}
for some constant \( c>0 \) and all \( v \) satisfying \( \| v \|\leq U_{n}. \)
On the other hand, Proposition~\ref{ExpBounds} (take \( G(u,z)=\exp(\i u^{\top} z) \))
implies
\begin{eqnarray*}
\P(\bar {\mathcal{A}})\leq B'n^{-1-\delta'}, \quad n\to \infty
\end{eqnarray*}
for some \( B'>0 \) and \( \delta'>0, \) provided \( D \) is large enough.
Therefore it holds on \( \mathcal{A} \)
\begin{eqnarray*}
     w(\| v \|)\left| \Psi_{\pi,n}(v)-\Psi_{\pi}(v) \right|  \leq C\sqrt{\frac{\log n}{n}}|\phi_{\pi }(v)|^{-1}\int_{[-1,1]^{m}}|1-\phi_{\pi}(v)/\phi_{\pi}(v+w)|\, dw
\end{eqnarray*}
for some constant \( C>0 \) and all \( v \) satisfying \( \| v \|\leq U_{n}. \) The rest of the proof is similar to the proof of Theorem~\ref{UpperBounds1}.
\subsection{Proof of Proposition~\ref{lower_bounds}}
In order to prove minimax lower bounds we apply general results from
\citet{T}. Let \( \Theta \) be a semi-parametric class of models.
Consider a family \( \{ P_{\theta}, \theta\in \Theta \} \) of probability measures, indexed by \( \Theta \).
For any \( \theta_{1}, \, \theta_{2}\in \Theta  \) let \( d(\theta_{1},\theta_{2}) \) be a semi-distance between two models
\( \theta_{1}  \) and \( \theta_{2}.  \)
\begin{lem}
\label{LB}
Suppose that \( \Theta \) contains two elements \( \theta_{1}  \) and \( \theta_{2}  \) such that
\( d(\theta_{1},\theta_{2})>2s \) for some \( s>0 \)
and \( \chi^{2}(P^{\otimes n}_{\theta_{1}},P^{\otimes n}_{\theta_{2}})\leq \tau <1/2, \) where
\[
\chi^{2}(P,Q)=:
\begin{cases}
\int\left( \frac{dP}{dQ}-1 \right)^{2}\,dQ, & \mbox{ if } P\ll Q \\
+\infty, & \mbox{otherwise}
\end{cases}
\]
for any two measures \( P \) and \( Q \). Then
\begin{eqnarray*}
    \inf_{\widehat \theta } \sup_{\theta\in \Theta}\P_{\theta}(d(\widehat \theta,\theta)\geq s )\geq c(\tau  )>0,
\end{eqnarray*}
where \( c(\tau) \) is constant depending on \( \tau   \) and infimum is taken over all estimates \( \widehat \theta  \) of \( \theta \) based on \( N \) observations under
\( P_{\theta} \).
\end{lem}
Turn now to the construction of models \( \theta_{1} \) and \( \theta_{2} \) from the class
\( \mathcal{A}(\bar \sigma, \varkappa)  \).
Let us consider a symmetric stable L\'evy model with a nonzero diffusion part (\( \sigma >0 \))
\[
  \psi(u)=\i\mu u-\sigma^{2}u^{2}/2+\vartheta(u), \quad \vartheta(u)=-\eta|u|^{\alpha},\quad 0<\alpha\leq 1, \quad u\in
  \mathbb{R}.
\]
For any \( \delta \) satisfying \( 0<\delta<\alpha \)  and \( M>0 \) define
\[
 \psi_{\delta}(u)=\i\mu u-\sigma^{2}u^{2}/2+\vartheta_{\delta}(u),
\]
with
\[
\vartheta_{\delta}(u)=-\eta|u|^{\alpha}\mathbf{1}_{\{ |u|\leq M \}}-
 \eta M^{\delta} |u|^{\alpha-\delta}\mathbf{1}_{\{ |u|>M
 \}}.
\]
Then \( \phi_{\delta}(u)=\exp(\psi_{\delta}(u))\) is the characteristic function  of some L\'evy process
and
\[
\phi_{\delta}(u)=\phi(u),\quad |u|\leq M,
\]
where \( \phi(u)=\exp( \psi(u)) \). Indeed, \( \vartheta_{\delta}(u) \)
is continuous, non-positive, symmetric function  which is convex on \( \mathbb{R}_{+} \)
for large enough \( M \).
According to the well known P\'olya criteria (see e.g. \citet{Ush}), the function \( \exp(\xi\vartheta_{\delta}(u)) \) is the
c. f. of some absolutely continuous distribution for
any \( \xi>0 \). In particular, for any natural \( k \) the function
\( \exp(\vartheta_{\delta}(u)/k) \) is the c. f. of some
absolutely continuous distribution.
Hence, \( \exp(\vartheta_{\delta}(u)) \) is the c.f. of some infinitely divisible distribution.
Define now two affine (in fact, L\'evy) models \( \theta_{1} \) and \( \theta_{2} \) corresponding
to the c.f. characteristic exponents \( \psi  \) and \( \psi_{\delta },  \) respectively.
Let \( \nu_{\theta_{1}}  \) and \( \nu_{\theta_{2}} \) be the corresponding L\'evy measures.
It holds
\begin{eqnarray*}
  \chi^{2}(P^{\otimes n}_{\theta_{1}},P^{\otimes n}_{\theta_{2}})=
  n\chi^{2}(p_{\theta_{1}},p_{\theta_{2}})&=&n\int_{\mathbb{R}}
  \frac{|p_{\theta_{1}}(y)-p_{\theta_{2}}(y)|^{2}}{p_{\theta_{1}}(y)}\,dy,
\end{eqnarray*}
where \( p_{\theta_{1}} \) and \( p_{\theta_{2}} \) are densities corresponding to c.f.
\( \phi_{\theta_{1}} \) and \( \phi_{\theta_{2}} \) respectively.
Using the asymptotic inequality
\[
p_{\theta_{1}}(y)\gtrsim |y|^{-(\alpha+1)},\quad |y|\to \infty
\]
and the fact that the density of the stable law  does
not vanish on any compact set in \( \mathbb{R} \), we derive
\begin{eqnarray*}
 n\chi^{2}(p_{\theta_{1}},p_{\theta_{2}})&\leq&
 nC_{1}\int_{|y|\leq A}|p_{\theta_{1}}(y)-p_{\theta_{2}}(y)|^{2}\,dy
\\
 &&+
 nC_{2}\int_{|y|>A}|y|^{\alpha+1}|p_{\theta_{1}}(y)-p_{\theta_{2}}(y)|^{2}\,dy=
 nC_{1}I_{1}+nC_{2}I_{2}
\end{eqnarray*}
for large enough \( A>0 \) and some constants \( C_{1},C_{2}>0 \).
Parseval's identity implies
\begin{eqnarray*}
I_{1}&\leq&
\frac{1}{2\pi}\int_{\mathbb{R}}|\phi_{\theta_{1}}(u)-\phi_{\theta_{2}}(u)|^{2}\,du
\\
&\leq&
\frac{1}{2\pi}\int_{|u|>M}e^{-\sigma^{2}|u|^{2}}\,du\lesssim
M^{-1}e^{-\sigma^{2} M^{2}}, \quad M\to \infty,
\end{eqnarray*}
\begin{eqnarray*}
I_{2}&\leq&
\frac{1}{2\pi}\int_{|u|>M}|(\phi_{\theta_{1}}(u)-\phi_{\theta_{2}}(u))'|^{2}\,du
\\
&\lesssim &
\int_{|u|>M}|u|^{2}e^{-\sigma^{2}|u|^{2}}\,du\lesssim
Me^{-\sigma^{2} M^{2}},\quad M\to \infty.
\end{eqnarray*}
The choice \( M\asymp\left[ \sigma^{-2}\log \left(n
\log^{\beta} n\right)\right]^{1/2} \) with
 some \( \beta>0 \)
yields
\[
n\chi^{2}(p_{\theta_{1}},p_{\theta_{2}})<1/2
\]
for large enough \( n. \)
On the other hand,
\begin{multline*}
\bar\vartheta(u)-\bar\vartheta_{\delta }(u)=
-\eta \int_{-1}^{1}\left[ |u|^{\alpha}\mathbf{1}_{\{ |u|>M\}}-|u+w|^{\alpha}\mathbf{1}_{\{ |u+w|>M\}} \right]\, dw
\\
+\eta M^{\delta}\int_{-1}^{1}\left[ |u|^{\alpha-\delta}\mathbf{1}_{\{ |u|>M\}}-|u+w|^{\alpha-\delta}\mathbf{1}_{\{ |u+w|>M\}} \right]\, dw
\end{multline*}
with
\begin{eqnarray*}
    \bar\vartheta(u)&:=& \int_{-1}^{1}\left[ \vartheta(u)-\vartheta(u+w) \right]\, dw,
    \\
    \bar\vartheta_{\delta }(u)&:=& \int_{-1}^{1}\left[ \vartheta_{\delta }(u)-\vartheta_{\delta }(u+w) \right]\, dw.
\end{eqnarray*}
Using the  identity
\begin{multline}
\label{asymp_vartheta}
 \int_{-1}^{1}\left[ |u|^{\alpha}\mathbf{1}_{\{ |u|>M\}}-|u+w|^{\alpha}\mathbf{1}_{\{ |u+w|>M\}} \right]\, dw=
 \\
 |u|^{\alpha}\int_{-1}^{1}\left[ 1-|1+w/u|^{\alpha} \right]\, dw=2\sum_{k=1}^\infty {\alpha \choose 2k} \frac{|u|^{\alpha-2k}}{2k+1},
\end{multline}
which holds for any \( |u|>M+1 \) and \( M>1 \), we get
\begin{eqnarray*}
\left| \int_{\mathbb{R}}[\bar\vartheta(u)-\bar\vartheta_{\delta }(u)] \, du \right|\, du\gtrsim M^{\alpha -1}, \quad M\to \infty
\end{eqnarray*}
for any \( 0<\alpha<1.  \)
Denote
\begin{eqnarray*}
    \rho_{\theta_{1}}(x)&:=&\int_{\mathbb{R}}e^{\mathfrak{i}ux}\bar\vartheta(u)\, du=\left( 1-\frac{\sin x}{x} \right)\nu_{\theta_{1}}(x),
    \\
    \rho_{\theta_{2}}(x)&:=&\int_{\mathbb{R}}e^{\mathfrak{i}ux}\bar\vartheta_{\delta }(u)\, du=\left( 1-\frac{\sin x}{x} \right)\nu_{\theta_{2}}(x),
\end{eqnarray*}
then the Fourier inversion formula implies
\begin{eqnarray}
\label{dmetric}
    d(\theta_{1},\theta_{2})&:=&\sup_{x\in \mathbb{R}}|\rho_{\theta_{1}}(x)-\rho_{\theta_{2}}(x)|
    \\
\notag
    &\geq & \left| \int_{\mathbb{R}}[\bar\vartheta(u)-\bar\vartheta_{\delta }(u)] \, du\right|
    \gtrsim M^{\alpha-1}, \quad M\to \infty.
\end{eqnarray}
Asymptotic expansion \eqref{asymp_vartheta} shows that there is a constant \( C \) depending on \( \eta \) such that
\begin{eqnarray*}
    |u|^{2-\alpha}\bar\vartheta(u)\leq C, \quad |u|^{2-\alpha}\bar\vartheta_{\delta} (u)\leq C, \quad u\in \mathbb{R}.
\end{eqnarray*}
Hence, taking \( \bar\sigma^{2}=\sigma^{2}/2,   \) \( \varkappa=2-\alpha   \),
we conclude that both models \( \theta_{1}  \) and \( \theta_{2} \) are in the class \( \mathcal{A}(\bar\sigma, \varkappa)  \).

\section{Appendix}

\subsection{Regularity properties of affine processes}
The next lemma provides bounds on the growth  of the derivatives of the conditional characteristic function
of an affine process.
\begin{lem}
\label{RegAffine}
If for some natural \( k>0, \)  the L\'evy measure  \( \nu^{(0)} \) satisfies
\begin{eqnarray}
\label{NuReg}
    \int_{\{ \|x\|>1 \}}\|x\|^{k}\nu^{(0)}(dx)<\infty,
\end{eqnarray}
then functions \( \psi_{0}(u,s) \) and \( \psi_{1}(u,s) \) from the representation \eqref{phi} are in \( C^{k+1}(\mathbb{R}_{+}) \) as functions of \( s \). Moreover, for any fixed \( x\in \mathcal{D} \) and \( s\in [0,T] \)  the following estimates hold
\begin{eqnarray}
\label{PhiDerBound}
    \left| \frac{\partial^{\mathbf{m}+j} \phi(u|s,x)}{\partial x^{\mathbf{m}}\partial s^{j}} \right|\leq C\| u \|_{2}^{2(|\mathbf{m}|+j)}, \quad \| u \|_{2}>1, \quad  j=0,\ldots,k+1,
\end{eqnarray}
where \( \mathbf{m} \) is a multi index and \( C \)  is a positive constant depending on \( s \) and \( x \).
\end{lem}
\begin{proof}
The existence of derivatives in \( s \) up to order \( k+1 \) was proved in \citet{DFS} (Lemma 6.5).
As was also shown in \citet{DFS} (Section 7), for any fixed \( s \) and \( x \), the function \( \phi(u|s,x) \)
is a c.f. of some infinitely divisible distribution, implying that
\begin{eqnarray*}
    \log\phi(u|s,x)=-u^{\top}Au+\i (b,u)+c+\int_{\mathcal{D}\setminus \{ 0 \}}(e^{\i y^{\top}u}-1-(\chi(y),u))M(dy),
\end{eqnarray*}
where \( A\in \mathbb{R}^{d\times d}, \) \( b\in \mathbb{R}^{d}, \) \( c\in \mathbb{R} \) and \( M \) is non-negative Borel measure on \( \mathcal{D}\setminus \{ 0 \}. \) It remains to note that \( A, \) \( b, \) \( c \) and \( M \) depend linearly on  \( x \) and are smooth in \( s, \) provided the  condition \eqref{NuReg} holds.
\end{proof}


\begin{thebibliography}{99}


\bibitem[A\"{i}t-Sahalia and Jacod(2009)]{AJ1} A\"{i}t-Sahalia, Y. and Jacod, J. (2006).
Volatility estimators for discretely sampled L\'evy processes. \textit{Annals of
Statistics}, \textbf{37}, 184--222.

\bibitem[A\"{i}t-Sahalia and Jacod(2008)]{AJ2} A\"{i}t-Sahalia, Y. and Jacod, J. (2009).
Estimating the degree of activity of jumps in high frequency financial data. \textit{Annals of
Statistics}, \textbf{37}(5A), 2202--2244.

\bibitem [Andreou, Ghysels and Kourtellos(2010)]{GAK}  Andreou, E., Ghysels, E. and Kourtellos, A. (2010). Forecasting with mixed-frequency data.  Oxford Handbook on Economic Forecasting edited by Michael P. Clements and David F. Hendry.


\bibitem [Belomestny(2009)]{Bel} Belomestny, D. (2009).  Spectral estimation of the fractional order of a L\'evy process. \textit{Annals of Statistics}, \textbf{38}(1), 317--351.

\bibitem[Bates(2005)]{B1}  Bates, D. (2000). Post-'87 crash fears in the S\&P 500 futures option market. \textit{Journal
of Econometrics}, \textbf{94}, 181--238.

\bibitem[Bates(2005)]{B2} Bates, D. (2005). Maximum Likelihood Estimation of
Latent Affine Processes. \textit{Review of Financial Studies}, 909--965.


\bibitem[Basawa and Brockwell(1982)]{BB} Basawa, I. V. and Brockwell, P. J. (1982).
Nonparametric estimation for nondecreasing L\'evy processes, \textit{J. Roy. Statist. Soc. Ser. B},
\textbf{44}, 262--269.


\bibitem[Cont and Mancini(2007)]{CM} Cont, R., and  Mancini C. (2004). Nonparametric Tests for Analyzing
 the Fine Structure of Price Fluctuations, SSRN Paper.


\bibitem[Duffie, Pan and Singleton(2000)]{DPS}
Duffie, D., Pan, J. and Singleton, K. (2000).
Transform analysis and asset pricing for affine jump diffusions.
\textit{Econometrica}, \textbf{68}, 1343--1376.

\bibitem[Duffie, Filipovi\'c and Schachermayer(2003)]{DFS}
Duffie, D., Filipovi\'c, D. and Schachermayer, W. (2003).
Affine processes and applications in finance.
\textit{Annals of Applied Prob.}, \textbf{13}, 984--1053.

\bibitem[Figueroa-L\'opez(2004)]{FL} Figueroa-L\'opez, J.E. (2004). Nonparametric estimation of L\'evy processes with a view towards mathematical finance. PhD thesis, Georgia Institute of Technology,
http://etd.gatech.edu. No. etd-04072004-122020.

\bibitem[Figueroa-L\'opez(2009)]{FL1}  Figueroa-Lopez, J.E.  (2009). Nonparametric estimation of time-changed Levy models under high-frequency data. To appear Advances in Applied Probability.


\bibitem[Glasserman and Kim(2007)]{GK} Glasserman, P. and Kyoung-kuk Kim (2007).
Moment Explosions and Stationary Distributions in Affine Diffusion Models,
to appear in \textit{Mathematical Finance}.


\bibitem[Jiang and Oomen(2007)]{JO} Jiang, G. and Oomen, R. (2007). Estimating Latent
variables and jump diffusion models using high-frequency data. \textit{Journal of Financial
Econometrics}, \textbf{5}, 1--30.

\bibitem[Jongbloed, van der Meulen and van der Vaart(2005)]{JMV}  Jongbloed, G., van der Meulen, F.H. and van der Vaart, A.W. (2005). Nonparametric inference
for L\'evy-driven Ornstein-Uhlenbeck processes. \textit{Bernoulli}, \textbf{11}(5), 759--791.





\bibitem[Keller-Ressel, Schachermayer  and Teichmann(2008)]{KST}  Keller-Ressel, M.,
Schachermayer, W. and  Teichmann, J.  (2008).  Affine processes are regular
, to appear in \textit{Probabilty Theory and Related Fields}.

\bibitem[Masuda(2007)]{MH} Masuda, H. (2007).
Ergodicity and exponential \( \beta \)-mixing bounds for multidimensional diffusions with jumps, \textit{Stochastic Process. Appl.}, \textbf{117}(1), 35-56.
\bibitem[Merlev\`ede, Peligrad and Rio(2009)]{MPR}  Merlev\'ede, F.,  Peligrad, M. and Rio, E. (2009). Bernstein inequality and moderate deviation under
strong mixing conditions. Working paper.
\bibitem[Neumann and Rei\ss(2007)]{NR} Neumann, M. and Rei\ss, M. (2007). Nonparametric
estimation for L\'evy processes from low-frequency observations. \textit{Bernoulli}, \textbf{15}(1), 223-248.
\bibitem[Rubin and Tucker(1959)]{RT} Rubin, H. and Tucker, H.G. (1959). Estimating the parameters
of a differential process, \textit{Ann. Math. Statist.}, \textbf{30}, 641-658.
\bibitem[Sato(1999)]{SA} Sato, K. (1999). \textit{L\'evy Processes and Infinitely Divisible Distributions.}
Cambridge University Press.
\bibitem[Singleton(2000)]{S} Singleton, K. (2001). Estimation of Affine Asset
Pricing Models Using the Empirical Characteristic Function.  \textit{Journal of Econometrics},  \textbf{10}, 111--141.
\bibitem[Stone(1982)]{St} Stone, C.J. (1982). Optimal global rates of convergence for nonparametric regression. \textit{Annals
of Statistics}, \textbf{10}, 1040--1053.
\bibitem [Tao, Wang, Yao and Zou(2010)]{TWYZ} Tao, M., Wang, Y., Yao, Q. and Zou, J. (2010). Large volatility matrix inference via combining low-frequency and high-frequency approaches. Technical report.
\bibitem[Tsybakov(2008)]{T} Tsybakov, A. (2008). \emph{Introduction to
Nonparametric Estimation}, Springer Series in Statistics, Springer.
\bibitem[Ushakov(1999)]{Ush} Ushakov, N. (1999). Selected topics in characteristic functions.
Modern Probability and Statistics. VSP, Utrecht.
\bibitem[Yokoyama(1980)]{Y}  Yokoyama, R. (1980). Moment bounds for stationary mixing sequences. \textit{Z. Wahrsch. Verw. Gebiete.}, \textbf{52}, 45--57.
\end{thebibliography}
\end{document}